\documentclass[aps,pra,superscriptaddress,twocolumn,10pt]{revtex4-1}

\usepackage{graphicx}
\usepackage{bm}
\usepackage[usenames,dvipsnames,svgnames,table]{xcolor}
\usepackage{epsfig}
\usepackage{amsmath}
\DeclareMathOperator{\Tr}{Tr}

\usepackage{amssymb}
\usepackage{array}
\usepackage{nccmath}
\usepackage{dsfont}
\usepackage{array}
\usepackage{cancel}
\usepackage{hyperref}
\hypersetup{colorlinks=true,breaklinks,linkcolor=blue,urlcolor=blue,citecolor=blue}

\usepackage{ulem}

\allowdisplaybreaks

\begin{document}
\title{Two-site reduced density matrix from one- and two-particle Green's functions}
\author{Gerg\H o Ro\'osz} 
\affiliation{%
    HUN-REN Wigner Research Centre for Physics, H-1525 Budapest, P.O.Box 49, Hungary}%
\author{Anna Kauch}
\affiliation{%
    Institute of Solid State Physics, TU Wien, 1040 Vienna, Austria}%
\author{Frederic Bippus}
\affiliation{%
    Institute of Solid State Physics, TU Wien, 1040 Vienna, Austria}%
\author{Daniel Wieser}
\affiliation{%
    Institute of Solid State Physics, TU Wien, 1040 Vienna, Austria}%
\author{Karsten Held}
\affiliation{%
    Institute of Solid State Physics, TU Wien, 1040 Vienna, Austria}%
    
\begin{abstract}
Strongly correlated electron systems 
 are challenging to calculate, and entanglement in such systems is not widely analyzed. 
 We present an approach that can be used as a post-processing step for calculating the two-site reduced density matrix and from it entanglement measures such as the mutual information and entanglement negativity. Input is only the one- and two-particle Green's function which is the output of numerous many-body methods.
 As an illustration, we present results for a toy model, the Hubbard model on a $2\times2$ cluster and a $6$ site ring.
\end{abstract}
\maketitle

There is recent interest in the entanglement properties of strongly correlated many-body systems in general \cite{Bayat2010, Bedoor2016, Erdmenger2016, Assaad2019, Assaad2018} and in the entanglement properties of the  Hubbard model in particular \cite{Walsh2019,Ehlers2015}. With the augmented perspective of entanglement, there is the  hope  of arriving at a better understanding of phase transitions and the complex physics of correlated electron systems.

In general, entanglement measures such as the von Neumann \cite{srednicki1993} or R\'enyi entropy \cite{Horodecki2009} at zero temperature or the mutual information \cite{Wolf2008} and entanglement negativity \cite{peres1996,HORODECKI1997}  at finite temperatures require knowing the reduced density matrix.
For wave function-based methods such as exact diagonalization (ED) \cite{Alvermann2011} or matrix product states (MPS) \cite{Cirac2021}, the reduced density matrix can be calculated directly.  The exact diagonalization is restricted to rather small fermionic systems; the MPS can overcome this exponential problem if the entanglement in the system is moderate as it is the case, e.g., in one dimension or in strip-like systems \cite{Walsh2019}. Then it is possible to calculate entanglement measures of extended subsystems.

For Green's function-based methods, on the other hand, or in experiments, we obtain typically only one- and two-particle Green's functions or measure corresponding spectral functions.
Determining the full reduced density matrix for an extensive subsystem is then impossible.
Green's function-based methods such as the dynamical vertex approximation (D$\Gamma$A) \cite{Rohringer2015} or (cluster) dynamical mean field theory \cite{GullRMP,Walsh2019} allow the simulation of larger systems and wider parameter range than MPS, quantum Monte Carlo, or ED.


The goal of the present paper is  to show the possibility of calculating the two-site reduced density matrix using one- and two-particle Green's functions.
The information of this two-site density matrix is equivalent to obtaining the expectation values of every hermitian operator, which can be defined on the two sites. Let us emphasize that the two sites can be at varying distances which allows us to extract much more information on entanglement than if the two sites were fixed. 


For qubits or spins, there is agreement on how to define useful measures of entanglement \cite{Eisert99, EisertPHD, Amico2008}. In comparison, entanglement for strongly correlated fermions is still in its infancy \cite{Amico2008,E:Sangchul2006,E:Hel13.1,E:Wag18.1,E:Toldin2018}. Among others, there is an ongoing
discussion on several types of entanglement measures \cite{Gigena2015,Ding2021,Ding2022,Ding2023,Ding2024}, the experimental usefulness of one-body fermionic entanglement  \cite{Krzysztof2023}, and the possible restriction (projection) of the reduced density matrix by the super-selection rule~\cite{Ding2021,Ding2022}.  The last is relevant if experiments are restricted to locally measurable operators.

As an illustration we calculate, in the present paper, the mutual information and the  entanglement negativity in small Hubbard clusters. The mutual information measures the total correlation between a bi-partition and is therefore an upper bound that the entanglement correlations cannot exceed. The negativity on the other hand can only be non-trivial if the system is entangled. Thus, a combination of both measures gives clear bounds on the parameter space for which entanglement is possible and provides a subspace for which entanglement certainly exists. Let us emphasize that having the reduced density matrix at hand, one can however calculate any entanglement or correlation measure based on it.

The mutual information is defined as follows: Let $A$  and $B$ be two selected sites of our much larger strongly correlated electron system that we consider. 
Then the reduced density matrix for both sites $A\cup B$ is $\rho_{A\cup B}$, and the reduced density matrices of the individual sites $A$ and $B$ are $\rho_A$ and $\rho_B$.  With these, the mutual information is defined as
\begin{align}
I&=  S_A + S_B - S_{A\cup B}\\
S_{A\cup B}&=-\Tr \rho_{A\cup B} \ln \rho_{A\cup B}  \\
S_A&=-\Tr \rho_A \ln \rho_A  \\
S_B&=-\Tr \rho_B \ln \rho_B  \; .
\end{align}

The entanglement negativity \cite{peres1996,HORODECKI1997} characterizes the entanglement between the two parts (here sites) of a system $A$ and $B$. To define it, we consider a basis $|n,m\rangle$ where $|n \rangle$ is a basis in the $A$ part, and $|m\rangle$ is a basis in the $B$ part. One defines the partial transpose of an operator $O$ in the Hilbert space of $A\cup B$
as 
\begin{equation}
	\langle n,m|O^{T_A}| r,s  \rangle \equiv \langle r,m|O| n,s  \rangle 
\end{equation}
The density matrix has only nonnegative eigenvalues; however in the case of entanglement between $A$ and $B$ the partial transpose $\rho^{T_A}$ of the density matrix may have negative eigenvalues. The negativity is defined as the sum of these negative eigenvalues
\begin{equation}
	{ N} = \sum_{\lambda_k<0} | \lambda_k| \; ,
\end{equation}   
where $\lambda_k$ denotes the spectrum of the partial transpose of the density matrix.

The outline of the paper is as follows: In Sec.~\ref{sec:model}, we introduce the Hubbard model to set the notation and units. In Sec.~\ref{sec:2siterho} the two-site reduced density matrix is defined and its elements explicitly written out. The equations of motion are used to reexpress 6- and 8-operator expectation values with 4-operator ones. The explicit expressions for the matrix elements of the density matrix in terms of one- and two-particle Green's functions are given in Sec.~\ref{sec:rhofromGF}, with additional derivations and reformulations given in the Appendices.
The results for mutual information and entanglement negativity obtained from the previously derived expressions for the density matrix elements for the 2-site, $2\times2$-site, and 6-site Hubbard model are presented in Sec.~\ref{sec:results}. We finally conclude the paper with Sec.~\ref{sec:conclusions}, where we also give an outlook.

\section{Model}
\label{sec:model}
In this work, we focus on the single-band Hubbard model with nearest neighbor hopping, but our formalism can also be extended to more complicated hoppings and interactions and multi-orbitals. The Hubbard model is defined by the Hamiltonian
\begin{equation}
	H=-t \sum_{\langle i,j\rangle,\sigma} c^{\dagger}_{i\sigma} c^{\phantom{\dagger}}_{j\sigma} + U \sum_i n_{i\uparrow} n_{i\downarrow} - \mu \sum_{i,\sigma} n_{i\sigma}, 
 \label{eq:Hubbard}
\end{equation}
where $c^{(\dagger)}_{i\sigma}$ is the fermion annihilation (creation) operator on-site $i$ with spin $\sigma$, $t$ is the hopping amplitude between neighboring sites on the lattice (the notation $\langle i,j\rangle$ indicates that the sum is only over pairs of nearest-neighboring sites $ i,j$ and no pair is counted twice), $U$ is the onsite Coulomb repulsion, $\mu$ is the chemical potential, and $n_{i\sigma} = c^{\dagger}_{i\sigma} c_{i\sigma}^{\phantom{\dagger}}$. In the following, we set $t\equiv 1$ as the unit of energy, frequency, and temperature (i.e. we also set $\hbar\equiv 1$ and $k_B \equiv 1$). 

The theoretical formalism presented in the paper is applicable to systems with arbitrary dimension. In the examples that we give in Sec.~\ref{sec:results} we focus on small size clusters (two, four, and six lattice sites) with periodic boundary conditions. In Sec.~\ref{sec:results} we present results both for the ground state ($T=0$) and finite temperature. We often use $\beta$ to denote the inverse of the temperature, i.e. $\beta\equiv 1/T$.

\section{Density matrix of two sites}
\label{sec:2siterho}
The basic quantity we consider in this paper is the reduced density matrix for two lattice sites. The (reduced) density matrix of any system can be written as \cite{Fagotti_2010} 
\begin{equation}
\rho=\sum_{i=1}^{d^2} A_i \langle A_i \rangle 
\label{eq:density_matrix}
\end{equation}
where $d$ is the dimension of the Hilbert space and $A_i$ for $i=1 \dots d^2$ forms a  basis in the space of the self-adjoint operators. 
\begin{align}
	A_i &= A^{\dagger}_i \\
	\Tr A_i A_j &= \delta_{i,j}
\end{align}
A well-known example of such a basis is the three (normalized) Pauli matrices plus the unit matrix for $d=2$. 
We will use  Eq.~(\ref{eq:density_matrix}) to construct the density matrix of the two sites, so we have to compute the expectation value of all operators.
To do so, one has to choose a basis, and the order of the basis vectors.

\begin{table}[tb]
	\begin{center}
		\begin{tabular}{|c |c c  c c |} 
			 & i$\uparrow$ & i$\downarrow$  & j$\uparrow$  & j$\downarrow$  \\ \hline
		   $v_1$ & 0 & 0  & 0  & 0 \\ \hline
		   $v_2$ & 0 & 0  & 0  & 1 \\
		   $v_3$ & 0 & 0  & 1  & 0 \\
		   $v_4$ & 0 & 1  & 0  & 0 \\
		   $v_5$ & 1 & 0  & 0  & 0 \\ \hline
		   $v_6$ & 0 & 0  & 1  & 1 \\
		   $v_7$ & 0 & 1  & 0  & 1 \\
		   $v_8$ & 1 & 0  & 0  & 1 \\ \hline
		   $v_9$ &  0 & 1  & 1  & 0 \\
		   $v_{10}$ & 1 & 0  & 1  & 0 \\
		   $v_{11}$ & 1 & 1  & 0  & 0 \\ \hline
		   $v_{12}$ & 0 & 1  & 1  & 1 \\
		   $v_{13}$ & 1 & 0  & 1  & 1 \\
		   $v_{14}$ & 1 & 1  & 0  & 1 \\
		   $v_{15}$ & 1 & 1  & 1  & 0 \\ \hline
		   $v_{16}$ & 1 & 1  & 1  & 1  \\ \hline
		\end{tabular}
	\end{center}
		\caption{Occupation number basis in the subspace of the two sites $i$ and $j$. The ith basis vector is mapped to the (16-i+1)th by the particle-hole transformation (so the first to the last and so on.) }
		\label{tab:check}
\end{table}
In the space of the self-adjoint operators, we choose the following basis:

\begin{align}
	A_{n,m}&=\frac{1}{\sqrt{2+2\delta_{n,m}}} ( |v_n  \rangle \langle  v_m| + | v_m \rangle \langle v_n | ) \\
	A^{\textrm{imag}}_{n,m}&=\frac{i}{\sqrt{2}} ( |v_n  \rangle \langle  v_m| - | v_m \rangle \langle v_n | ) 
\end{align}
	
Here $A_{n,m}$ for $n=m$ contains the operators that project onto a basis vector $| v_{n=m} \rangle$; for $n \neq m$  it is an operator which mixes the two basis vectors and has real components \footnote{This changes if the Hamiltonian becomes complex, e.g., when applying a magnetic field. In such a case, $A_{n,m}$ has real and imaginary parts.}. For every two basis vectors there is a mixing operator with a purely imaginary matrix, these are the $A^{\textrm{imag}}_{n,m}$ operators. All $A_{n,m}$ and all $A^{\textrm{imag}}_{n,m}$ together form the basis in the operator space.

Our Hamiltonian is a real matrix on the occupation number basis, so the eigenvectors are also real. 
 To fullfill this requirement all $\langle A^{\textrm{imag}}_{n,m} \rangle$ must be purely imaginary numbers which contradicts the requirement that expectation values of self-adjoint operators must be real, therefore \begin{equation}
		\langle A^{\textrm{imag}}_{n,m}\rangle =0 \textrm{ for} \;  n,m=1 \dots 16.
\end{equation} 
There is spin conservation and also particle number conservation. With these symmetries, the density matrix has only 26 non-zero elements (of $16\times16=256$). (Here the particle-hole symmetry is not used.) The structure of the density matrix in this basis can be seen in Eq.~(\ref{eq:matrix_of_density_matrix}). All elements that are not displayed are zero.

\begin{widetext}
\begin{equation}
	\newcommand*{\ltemp}{\multicolumn{1}{c|}{}}
	\newcommand*{\rtemp}{\multicolumn{1}{|c}{}}
\label{eq:matrix_of_density_matrix}
	\rho=\left[\begin{array}{cccccccccccccccc}
		\rho_{1,1} & \rtemp & &  & & & & & & & & & & & & \\ \cline{1-5}
		 \ltemp & \rho_{2,2} &  & \rho_{2,4} &  & \rtemp & & & & & & & & & &   \\
		 \ltemp &  & \rho_{3,3} &  & \rho_{3,5} & \rtemp & & & & & & & & & &   \\
		 \ltemp & \rho_{2,4} &  &  \rho_{4,4}&  & \rtemp & & & & & & & & & &   \\
		 \ltemp &  & \rho_{3,5} &  & \rho_{5,5} & \rtemp & & & & & & & & & &   \\ \cline{2-11}
		  &  &  &  & \ltemp & \rho_{6,6} &  & \rho_{6,8} & \rho_{6,9} &  & \rho_{6,11} & \rtemp  & & & & \\
		  &  &  &  & \ltemp &  & \rho_{7,7} & & & & & \rtemp & & & & \\
		  &  &  &  & \ltemp & \rho_{6,8} & & \rho_{8,8} & \rho_{8,9} & & \rho_{8,11}& \rtemp & & & & \\
		  &  &  &  & \ltemp & \rho_{6,9} & &\rho_{8,9} & \rho_{9,9} & & \rho_{9,11} & \rtemp & & & & \\
		  &  &  &  & \ltemp &  & & & & \rho_{10,10} & & \rtemp & & & & \\
		  &  &  &  & \ltemp &  \rho_{6,11} &  & \rho_{8,11} & \rho_{9,11} & & \rho_{11,11} & \rtemp & & & & \\ \cline{6-15} 		    
		  &  &  &  &  &   &  &  & & & \ltemp & \rho_{12,12} & & \rho_{12,14} & & \rtemp \\
		  &  &  &  &  &   &  &  & & & \ltemp &  & \rho_{13,13}& & \rho_{13,15} & \rtemp \\
		  &  &  &  &  &   &  &  & & & \ltemp & \rho_{12,14} & & \rho_{14,14}& & \rtemp \\
		  &  &  &  &  &   &  &  & & & \ltemp &  & \rho_{13,15} &  &  \rho_{15,15}& \rtemp \\ \cline{12-16}
		  &  &  &  &  &   &  &  & & &  &  & & & \ltemp & \rho_{16,16}  \\ 
	\end{array}\right]
\end{equation}
\end{widetext}
 For every matrix element of $\rho$ we have $\rho_{n,m}=\langle A_{n,m}\rangle$, with the 
   diagonal operators $A_{n,n}$ being (in second quantized form)
   

\begin{align}
	A_{1,1} &= (1-n_{i\uparrow})&(1-n_{i\downarrow})&(1-n_{j\uparrow})&(1-n_{j\downarrow}) \label{eq:diag}\\
	A_{2,2} &= (1-n_{i\uparrow})&(1-n_{i\downarrow})&(1-n_{j\uparrow})&n_{j\downarrow} \\
	 A_{3,3} &= (1-n_{i\uparrow})&(1-n_{i\downarrow})&n_{j\uparrow}&(1-n_{j\downarrow}) \\
	A_{4,4} &= (1-n_{i\uparrow})&n_{i\downarrow}&(1-n_{j\uparrow})&(1-n_{j\downarrow}) \\
	 A_{5,5} &= n_{i\uparrow}&(1-n_{i\downarrow})&(1-n_{j\uparrow})&(1-n_{j\downarrow}) \\
	 A_{6,6} &= (1-n_{i\uparrow})&(1-n_{i\downarrow})&n_{j\uparrow}&n_{j\downarrow} \\
	 A_{7,7} &= (1-n_{i\uparrow})&n_{i\downarrow}&(1-n_{j\uparrow})&n_{j\downarrow} \\
	 A_{8,8} &= n_{i\uparrow} &(1-n_{i\downarrow})&(1-n_{j\uparrow})&n_{j\downarrow} \\
	 A_{9,9} &= (1-n_{i\uparrow})& n_{i\downarrow} & n_{j\uparrow}&(1-n_{j\downarrow}) \\
	 A_{10,10} &= n_{i\uparrow} & (1-n_{i\downarrow}) & n_{j\uparrow} &(1-n_{j\downarrow}) \\
	A_{11,11} &= n_{i\uparrow}&n_{i\downarrow}&(1-n_{j\uparrow})&(1-n_{j\downarrow}) \\
	A_{12,12} &= (1-n_{i\uparrow})&n_{i\downarrow}&n_{j\uparrow}&n_{j\downarrow} \\
	A_{13,13} &= n_{i\uparrow}&(1-n_{i\downarrow})&n_{j\uparrow}&n_{j\downarrow} \\
	A_{14,14} &= n_{i\uparrow}&n_{i\downarrow}&(1-n_{j\uparrow})&n_{j\downarrow} \\
	A_{15,15} &= n_{i\uparrow}&n_{i\downarrow}&n_{j\uparrow}&(1-n_{j\downarrow}) \\
	A_{16,16} &= n_{i\uparrow}&n_{i\downarrow}&n_{j\uparrow}&n_{j\downarrow} 
\end{align}

For the non-diagonals, we get 
\begin{align}
A_{2,4}&=\frac{1}{\sqrt{2}} ( c^{\dagger}_{i\downarrow}c_{j \downarrow}+c^{\dagger}_{j\downarrow}c_{i\downarrow})(1-n_{i\uparrow})(1-n_{j,\uparrow}) \\
A_{3,5}&=\frac{1}{\sqrt{2}} ( c^{\dagger}_{i\uparrow}c_{j\uparrow}+c^{\dagger}_{j\uparrow}c_{i\uparrow})(1-n_{i\downarrow})(1-n_{j,\downarrow}) \\
A_{8,6}&=\frac{1}{\sqrt{2}} ( c^{\dagger}_{i\uparrow}c_{j\uparrow}+c^{\dagger}_{j\uparrow}c_{i\uparrow})(1-n_{i\downarrow})n_{j,\downarrow} \\
A_{9,6}&=-\frac{1}{\sqrt{2}} ( c^{\dagger}_{i\uparrow}c_{j\uparrow}+c^{\dagger}_{j\uparrow}c_{i\uparrow})(1-n_{i\uparrow}) n_{j,\uparrow}  \\
A_{11,6}&=\frac{1}{\sqrt{2}} (  c^{\dagger}_{i\uparrow}c^{\dagger}_{i \downarrow} c_{j\downarrow}c_{j\uparrow} 
 +  c^{\dagger}_{j\uparrow} c^{\dagger}_{j\downarrow} c_{i \downarrow} c_{i\uparrow}) \\
 A_{8,9}&=\frac{1}{\sqrt{2}} (  c^{\dagger}_{j\uparrow}c^{\dagger}_{i \downarrow} c_{i\uparrow}c_{j\downarrow} 
 +  c^{\dagger}_{j\downarrow} c^{\dagger}_{i\uparrow} c_{i \downarrow} c_{i\uparrow}) \\
 A_{8,11}&=\frac{1}{\sqrt{2}} ( c^{\dagger}_{j\downarrow}c_{i,\downarrow}+c^{\dagger}_{i\downarrow}c_{j\downarrow})n_{i\uparrow}(1-n_{j,\uparrow}) \\
 A_{9,11}&=\frac{-1}{\sqrt{2}} ( c^{\dagger}_{j\uparrow}c_{i,\uparrow}+c^{\dagger}_{i\uparrow}c_{j\uparrow})n_{i\downarrow}(1-n_{j,\downarrow}) \\
 A_{12,14}&=\frac{-1}{\sqrt{2}} ( c^{\dagger}_{j\uparrow}c_{i\uparrow}+c^{\dagger}_{i\uparrow}c_{j\uparrow})n_{i\downarrow}n_{j,\downarrow} \\
 A_{13,15}&=\frac{-1}{\sqrt{2}} ( c^{\dagger}_{j\downarrow}c_{i\downarrow}+c^{\dagger}_{i\downarrow}c_{j\downarrow})n_{i\uparrow}n_{j\uparrow} 
\label{eq:non-diag}
 \end{align}

The expectation values of all these operators will now be written in terms of four-point and two-point functions and the derivatives of the four-point functions.
Actually, some of the operators in Eqs.~\eqref{eq:diag}-~\eqref{eq:non-diag} are already expressed by four or fewer fermion operators, and therefore easy to connect to four-point functions (when taking the expectation value).  However others contain terms with 6 and 8 fermion operators.
To calculate these we use the equations of motion in imaginary time, and relate the 6 and 8 operator terms to single and double imaginary time derivatives of the correlators in the following.

This is a formidable task \footnote{For spin systems, it is much easier and the  $4\times 4$ reduced density matrix can be related to the equal-time spin-spin correlation functions \cite{Amico2004}.}, and we start with
the equation of motion for the annihilation operator  $c_{i\uparrow}$ which reads for the Hubbard model with nearest neighbor hopping
\begin{align}
	\frac{\partial c_{i\uparrow}(\tau)}{\partial \tau} = [H,c_{i\uparrow}]&=t \sum_{\delta}c_{i+\delta\uparrow} - \mu c_{i\uparrow}- U c_{i \uparrow} c^{\dagger}_{i\downarrow} c_{i\downarrow},.
\end{align}
Here $\delta$ sums over all nearest neighbors $i+\delta$ of site $i$. In principle, we would have another set of $\delta'$'s for every $i$, but for the sake of compactness we omit a subindex. In a numerical implementation $i$ is typically  a $d$-dimensional tuple in terms of the basis vectors of the lattice. For example, for the  square lattice we would add $\delta \in\{(1,0), (-1,0), (0,1), (0,-1)\}$ to
$i=(i_x,i_y)$ (in this notation $\delta$ is then anyhow independent of $i$).
 
To get the aforementioned 8-point function $ \langle n_{i\uparrow} n_{i\downarrow} n_{j\uparrow} n_{j\downarrow} \rangle$, one can consider the derivative 
$\langle  \frac{\partial}{\partial \tau_1} c_{i\uparrow}(\tau_1) c^{\dagger}_{i\uparrow}(\tau_2) \frac{\partial}{\partial \tau_3} c_{j\uparrow}(\tau_3) c^{\dagger}_{j\uparrow}(\tau_4) \rangle $ and
either calculate it directly or by relating it to the derivatives of two-particle Green's functions (as we will do later). On the other hand, using the equation of motion, we can express the double derivative as a sum of the 8-point function and  6 and 4 fermionic operator expectation values. For the 6-point expectation values one can use  the equation of motion again, and then collect the terms. Some terms cancel each other and the final expression contains maximally 4-point correlators. Let us consider explicitly the following term, which contributes to  $\langle A_{7,7} \rangle$: 
\begin{align}
  \langle c_{i\uparrow} (\tau_1) &c^{\dagger}_{i\downarrow} (\tau_1) c_{i\downarrow} (\tau_1)  c^{\dagger}_{i \uparrow}(\tau_2) c_{j\uparrow} (\tau_3)c^{\dagger}_{j\downarrow} (\tau_3) c_{j\downarrow} (\tau_3)  c_{j\uparrow}^{\dagger}(\tau_4)  \rangle \nonumber \\
 =& \frac{1}{U^2} \frac{\partial }{\partial \tau_1} \frac{\partial}{\partial \tau_3} \langle  c_{i\uparrow}(\tau_1) c^{\dagger}_{i\uparrow}(\tau_2) c_{j\uparrow} (\tau_3) c^{\dagger}_{j\uparrow} (\tau_4)\rangle 
 	\nonumber\\
 - \frac{t}{U^2} &\sum_{\delta} \left[   \frac{\partial}{\partial \tau_3} \langle  c_{i+\delta\uparrow}(\tau_1) c^{\dagger}_{i\uparrow}(\tau_2) c_{j\uparrow} (\tau_3) c^{\dagger}_{j\uparrow} (\tau_4)\rangle \right. \nonumber\\  
  +& \left. \frac{\partial }{\partial \tau_1}  \langle  c_{i\uparrow}(\tau_1) c^{\dagger}_{i\uparrow}(\tau_2) c_{j+\delta\uparrow} (\tau_3) c^{\dagger}_{j\uparrow} (\tau_4)\rangle \right] \nonumber\\
   -&\frac{\mu}{U^2} \frac{\partial}{\partial \tau_3} \langle  c_{i\uparrow}(\tau_1) c^{\dagger}_{i\uparrow}(\tau_2) c_{j\uparrow} (\tau_3) c^{\dagger}_{j\uparrow} (\tau_4)\rangle \nonumber\\
   -&\frac{\mu}{U^2} \frac{\partial}{\partial \tau_1} \langle  c_{i\uparrow}(\tau_1) c^{\dagger}_{i\uparrow}(\tau_2) c_{j\uparrow} (\tau_3) c^{\dagger}_{j\uparrow} (\tau_4)\rangle \nonumber\\
   +\frac{t^2}{U^2}& \sum_{\delta,\Delta}  \langle  c_{i+\delta\uparrow}(\tau_1) c^{\dagger}_{i\uparrow}(\tau_2) c_{j+\Delta\uparrow} (\tau_3) c^{\dagger}_{j\uparrow} (\tau_4)\rangle \nonumber\\
   + \frac{ \mu t}{U^2} &\sum_{\delta} \left[   \langle  c_{i+\delta\uparrow}(\tau_1) c^{\dagger}_{i\uparrow}(\tau_2) c_{j\uparrow} (\tau_3) c^{\dagger}_{j\uparrow} (\tau_4)\rangle \right. \nonumber\\ 
   +& \left.   \langle  c_{i\uparrow}(\tau_1) c^{\dagger}_{i\uparrow}(\tau_2) c_{j+\delta\uparrow} (\tau_3) c^{\dagger}_{j\uparrow} (\tau_4)\rangle \right] \nonumber\\
   +& \frac{\mu^2}{U^2} \langle  c_{i\uparrow}(\tau_1) c^{\dagger}_{i\uparrow}(\tau_2) c_{j\uparrow} (\tau_3) c^{\dagger}_{j\uparrow} (\tau_4)\rangle,
   \label{eq:sec_deriv}
\end{align}
with $\delta$ and $\Delta$ denoting all possible shifts to  nearest neighbors of the site indices $i$ and $j$, respectively. Please note, that we restrict ourselves to the case of nearest-neighbor hopping here. For longer range hoppings there would occur further terms from the commutator, but the overall calculation would be very similar. To obtain the final expression for the (equal time) density matrix elements, all four imaginary times are set to zero: $\tau_1\to\tau_2^+\to\tau_3^+\to\tau_4^+\to 0^+$.

There are two additional 6-operator terms needed for the density matrix. For the diagonal, a product of three different number operators is needed - due to the spin and site interchange symmetries it does not matter which three so let us thus consider only $\langle  n_{i\uparrow} n_{i\downarrow} n_{j \uparrow }\rangle$ in the following.
For the off-diagonals, we need, for similar symmetry arguments, only one kind of 6-operator term, say $\langle c^{\dagger}_i  c_j  n_{j\uparrow} n_{i\uparrow} \rangle $.

The equation of motion relating $\langle  n_{i\uparrow} n_{i\downarrow} n_{j \uparrow }\rangle$ to the four operator expectation values is
\begin{align}
&\langle n_{i\uparrow} n_{i\downarrow} n_{j \uparrow } \rangle = \langle n_{i\downarrow}\rangle - \langle n_{i\downarrow} n_{j\uparrow}\rangle - \langle n_{i\uparrow} n_{i \downarrow} \rangle \nonumber\\
& -\frac{t}{U} \sum_{\delta} \langle  c_{i+\delta \uparrow} c^\dagger_{i\uparrow} c_{j \uparrow} c^{\dagger}_{j \uparrow}\rangle   -\frac{\mu}{U}  \langle  c_{i \uparrow} c^\dagger_{i\uparrow} c_{j \uparrow} c^{\dagger}_{j \uparrow}\rangle \nonumber \\
& + \frac{1}{U} \frac{\partial}{\partial \tau_1}    \langle  c_{i \uparrow} (\tau_1) c^\dagger_{i\uparrow} (\tau_2) c_{j \uparrow} (\tau_3) c^{\dagger}_{j \uparrow} (\tau_4)\rangle, 
\end{align}
where the imaginary time dependence is only explicitly written out in the last term.
Similarly, the equation of motion used for the off-diagonal terms  is
\begin{align}
&\langle c^{\dagger}_{i\downarrow} c_{j \downarrow} n_{j\uparrow} n_{i\uparrow}\rangle =-\langle c_{j\downarrow} c^{\dagger}_{i\downarrow}\rangle +  \langle  c_{j\downarrow} c^\dagger_{i\downarrow} c_{j\uparrow} c^{\dagger}_{j\uparrow}\rangle \nonumber \\
& -\frac{t}{U} \sum_{\delta} \langle  c_{j+\delta \downarrow} c^\dagger_{i\downarrow} c_{i \uparrow} c^{\dagger}_{i \uparrow}\rangle + \frac{\mu}{U} \langle  c_{j \downarrow} c^\dagger_{i\downarrow} c_{i \uparrow} c^{\dagger}_{i \uparrow}\rangle \nonumber\\
&-\frac{1}{U} \frac{\partial}{\partial \tau_1}  \langle  c_{j \downarrow} c^\dagger_{i\downarrow} c_{i \uparrow} c^{\dagger}_{i \uparrow}\rangle.   
\end{align}
Using the spin conservation and the equivalence of the two sites, one finds that a lot of matrix elements are equal. 
To get the matrix elements from  correlators and derivatives of correlators, one uses the above-mentioned equations of motion and collects all terms.

After we have seen, how the higher-order correlators can be
expressed through time-derivatives of lower-order correlators, let us express all elements of the reduced two-site density matrix by a number of correlators ${\cal C}$, which we define later and which are combinations of the two- and four-point (one- and two-particle) Green's function.

The off-diagonal terms are given by the following expressions
\begin{align}
& \rho_{11,6}={\cal C}_{10}  \label{eq:rho116}\\
& \rho_{8,9}={\cal C}_{11} \\
 &\rho_{2,4}=\rho_{3,5} =\left(\frac{\mu}{U}-1\right){\cal C}_{8A}+\frac{t}{U} {\cal C}_{12} -\frac{1}{U} {\cal C}_4 \\
 &\rho_{8,6}=\rho_{8,11}=-\rho_{9,11}=-\rho_{9,6} \nonumber\\
 &=\left (\frac{1}{2}-\frac{\mu}{U}\right){\cal C}_{8A}-\frac{1}{2}{\cal C}_{8B}-\frac{t}{U}{\cal C}_{12}+ \frac{1}{U}{\cal C}_4 \\
&\rho_{12,14}=\rho_{13,15} ={\cal C}_9-{\cal C}_{8A}-\frac{t}{U} {\cal C}_{12} +  \frac{1}{U}{\cal C}_4 - \frac{\mu}{U}{\cal C}_{8A}
\end{align}
where the  two- and four-point correlators ${\cal C}$'s are given explicitly in the next section.

The $\rho_{7,7}$ matrix element is given by
\begin{align}
    &\rho_{7,7}=\rho_{10,10}= \nonumber\\
    &\frac{1}{U^2}{\cal C}_1 -\frac{t}{U^2} {\cal C}_t
-\frac{\mu}{U^2} {\cal C}_{\mu} +\frac{t^2}{U^2} {\cal C}_{t^2}
+\frac{\mu t}{U^2} {\cal C}_{\mu t}
+\frac{\mu^2}{U^2} {\cal C}_{5}
\label{eq:rho1010}
\end{align}
Before calculating the remaining diagonal matrix elements, we provide, as an intermediate step, expressions for the following expectation values
\begin{align}
    \langle n_{i\uparrow} \rangle &= 1 - {\cal C}_{13} \label{eq:niup} \\
    \langle n_{i\uparrow} n_{j \uparrow} \rangle &= -1+2 \langle n_{i\uparrow} \rangle+{\cal C}_5 \\
    \langle n_{i\uparrow} n_{j \downarrow} \rangle &= -1+2 \langle n_{i\uparrow} \rangle+{\cal C}_6 \\
    \langle n_{i\uparrow} n_{i \downarrow} \rangle &= -1+2 \langle n_{i\uparrow} \rangle+{\cal C}_7 \label{eq:niupnidown}\\
    \langle n_{i\uparrow} n_{i \downarrow} n_{j \uparrow}\rangle &= -\langle n_{i\uparrow} \rangle + \langle n_{i\uparrow} n_{j \downarrow} \rangle + \langle n_{i\uparrow} n_{i \downarrow} \rangle \nonumber\\
    &+\frac{t}{U} {\cal C}_{\mu t 1}+\frac{\mu}{U} {\cal C}_5 -\frac{1}{U} {\cal C}_{\mu 2} \\
    \langle n_{i\uparrow} n_{i \downarrow} n_{j \uparrow} n_{j \downarrow} \rangle &= 
    \rho_ {7,7}+2 \langle n_{i\uparrow} n_{i \downarrow} n_{j \uparrow}\rangle - \langle n_{i\uparrow} n_{j \uparrow} \rangle.
\end{align}
With these expectation values, the diagonal matrix elements can be calculated as
\begin{align}
    &\rho_{16,16}=\langle n_{i\uparrow} n_{i \downarrow} n_{j \uparrow} n_{j \downarrow} \rangle  \label{eq:rho1616} \\
    &\rho_{1,1}= \langle n_{i\uparrow} n_{i \downarrow} n_{j \uparrow} n_{j \downarrow} \rangle -4\langle n_{i\uparrow} n_{i \downarrow} n_{j \uparrow}\rangle+2\langle n_{i\uparrow} n_{i \downarrow} \rangle \nonumber \\
    &+ 2\langle n_{i\uparrow} n_{j \uparrow} \rangle +2\langle n_{i\uparrow} n_{j \downarrow} \rangle- 4 \langle n_{i\uparrow} \rangle +1 \\ 
    &\rho_{2,2}=\rho_{3,3}=\rho_{4,4} = \rho_{5,5} \nonumber\\
    &=-\langle n_{i\uparrow} n_{i \downarrow} n_{j \uparrow} n_{j \downarrow} \rangle + 3 \langle n_{i\uparrow} n_{i \downarrow} n_{j \uparrow}\rangle - \langle n_{i\uparrow} n_{j \uparrow} \rangle\nonumber  \\
    & - \langle n_{i\uparrow} n_{j \downarrow} \rangle -\langle n_{i\uparrow} n_{i \downarrow} \rangle + \langle n_{i\uparrow} \rangle \\
    &\rho_{6,6}=\rho_{11,11} \nonumber \\
    &=\langle n_{i\uparrow} n_{i \downarrow} n_{j \uparrow} n_{j \downarrow} \rangle-2\langle n_{i\uparrow} n_{i \downarrow} n_{j \uparrow}\rangle + \langle n_{i\uparrow} n_{i \downarrow} \rangle \\
      &\rho_{8,8}=\rho_{9,9} \nonumber \\
    &=\langle n_{i\uparrow} n_{i \downarrow} n_{j \uparrow} n_{j \downarrow} \rangle-2\langle n_{i\uparrow} n_{i \downarrow} n_{j \uparrow}\rangle + \langle n_{i\uparrow} n_{j \downarrow} \rangle \\
    &\rho_{12,12}=\rho_{13,13}=\rho_{14,14}=\rho_{15,15} \nonumber \\
    &=-\langle n_{i\uparrow} n_{i \downarrow} n_{j \uparrow} n_{j \downarrow} \rangle 
    +\langle n_{i\uparrow} n_{i \downarrow} n_{j \uparrow}\rangle.
     \label{eq:rho1212}
\end{align}
With Eq.~(\ref{eq:niup}) and  Eq.~(\ref{eq:niupnidown}), we can also calculate the reduces density matrix $\rho_{A(B)}$ of a single site $A$($B$) that is needed for the mutual information, see Appendix~\ref{Appendix1siteRD}. 
Further Appendix~\ref{AppendixEV} lists the eigenvalues of the two-site density matrix $\rho$ [Eq.~(\ref{eq:matrix_of_density_matrix})].

\section{Connection to the one- and two-particle Green's functions}
\label{sec:rhofromGF}

For calculating the reduced two-site density matrix with the equations of the last section, we still need the correlators $\cal C$. These can be calculated directly in the position and imaginary time variables. In the case of extended systems with translational symmetry, we can reexpress all of the correlators through one- and two-particle Green's functions dependent on momentum and Matsubara frequency. The derivatives with respect to imaginary time can be replaced by multiplication with the respective Matsubara frequency.  Please note, once again, that the equal time correlators occurring in the above expressions are not Green's functions, since they do not contain the time ordering operator $\cal{T}$ (they are already time ordered). By appropriate reordering of the operators, we can always obtain them as  Green's functions in the equal time limit. Special care has to be taken for reexpressing of time derivatives of already time-ordered correlators. 

\subsection{Derivatives of the Green's functions}
Before finally turning to the connection of the correlators ${\cal C}$ to Green's functions, it is helpful to explicitly look at the time derivative.
In the equal time limit, the derivative of the Green's function has a divergent term originating from the time-ordering operator: 
  \begin{align}
 &\frac{\partial }{\partial \tau_1}  G_{12}(\tau_1-\tau_2)  = -\frac{\partial }{\partial \tau_1} \langle T[c_1(\tau_1) c^{\dagger}_2(\tau_2] \rangle \nonumber \\
 & = -\delta(\tau_1 - \tau_2)\delta_{12} - \langle T [\frac{\partial }{\partial \tau_1}  c_1 (\tau_1) c^{\dagger}_2(\tau_2)] \rangle. 
  \end{align}
We will later need only the already time-ordered correlators $\langle \frac{\partial }{\partial \tau_1}  c_1 (\tau_1) c^{\dagger}_2(\tau_2) \rangle $ in the limit $\tau_1\to 0^+$, $\tau_2=0$, which in the Matsubara frequency representation gives the following expression:
\begin{align}
     &\left \langle\frac{\partial }{\partial \tau_1}  c_1 (\tau_1) c^{\dagger}_2 \right\rangle = 
     -\frac{\partial }{\partial \tau_1}  G_{12}(\tau_1) - \delta(\tau_1)\delta_{12}
     \nonumber \\
     &= \frac{1}{\beta}\sum_\nu i\nu \left [G_{12}(\nu) - \frac{1}{i\nu}\delta_{12} \right ] e^{-i\nu\tau_1}, \quad  \tau_1\to 0^+,
\end{align}
where 
$
G_{12}(\nu) = \int_0^\beta d \tau e^{i\nu\tau} G_{12}(\tau).
$ Similarly for the two-particle Green's function 
\begin{align}
 G_{1234}(\tau_1,\tau_2,\tau_3, \tau_4)  = 
 \langle T [  c_1(\tau_1) c^{\dagger}_2(\tau_2)  c_3 (\tau_3) c^{\dagger}_4 (\tau_4)]\rangle 
 \end{align}
 we need to express the time derivatives of the already time-ordered correlators as a difference of the Green's function and the Dirac-$\delta$ terms resulting from the derivative of the time-ordering operator. It turns out that in the equal time limit, we obtain the derivative of the connected part of the Green's functions plus disconnected terms. For example, the doubly derived correlator in the first term on the right-hand side of \eqref{eq:sec_deriv} that we need in $A_{7,7}$ takes the form 
\begin{align}
& \lim_{\tau_1\to\tau_2\to\tau_3\to\tau_4=0} \langle \frac{\partial }{\partial \tau_1}   c_1(\tau_1) c^{\dagger}_2(\tau_2) \frac{\partial }{\partial \tau_3}  c_3 (\tau_3) c^{\dagger}_4 (\tau_4)\rangle \nonumber \\
&= \lim_{\tau_1\to\tau_2\to\tau_3\to\tau_4=0} \left [\frac{\partial }{\partial \tau_1}\frac{\partial }{\partial \tau_3} G^{conn}_{1234} \right] \nonumber \\
& +\lim_{\tau_1\to0^+} \left [  \left \langle\frac{\partial }{\partial \tau_1}  c_1 (\tau_1) c^{\dagger}_2 \right\rangle \right] \lim_{\tau_3\to0^+}\left [ \left\langle\frac{\partial }{\partial \tau_3}  c_3  (\tau_3)c^{\dagger}_4 \right\rangle \right]  \nonumber \\
&-\lim_{\tau_1\to0^+} \left [  \left \langle\frac{\partial }{\partial \tau_1}  c_1 (\tau_1) c^{\dagger}_4 \right\rangle \right] \lim_{\tau_3\to0^-}\left [ \left\langle\frac{\partial }{\partial \tau_3}  c_3  (\tau_3)c^{\dagger}_2 \right\rangle \right]\nonumber \\
& \equiv  \lim_{\tau_1\to\tau_2\to\tau_3\to\tau_4=0} \left [ \frac{\partial }{\partial \tau_1}\frac{\partial }{\partial \tau_3} \widehat{{G}}_{1234}(\tau_1,\tau_2,\tau_3, \tau_4) \right]. 
\label{eq:gbarwave}
 \end{align}
 In the last line we denoted the modified Green's-function-like expression (connected part of the Green's function plus modified disconnected terms) as $\widehat{{G}}_{1234}$, which we can thus write (omitting time variables) as
 \begin{align}
 {\widehat{G}}_{1234}  = G^{conn}_{1234} + \langle c_1  c^{\dagger}_2 \rangle \langle c_3  c^{\dagger}_4 \rangle -\langle c_1  c^{\dagger}_4 \rangle \langle c_3  c^{\dagger}_2 \rangle .   
 \end{align}
 The explicit derivations and definitions of this term and also other modified Green's functions: ${\overline{G}}_{1234}$ and $\widetilde{{G}}_{1234}$ are given in the Appendix~\ref{sec:AppA}. 

 \subsection{Explicit expressions for the $\cal C$ correlators}
In the following expressions for the correlators ${\cal C}$, we use a combined notation $k=(\vec{k},\nu)$, $q=(\vec{q},\omega)$, where $\vec{k}$ and $\vec{q}$ are momenta and  $\nu$, $\omega$ are discrete Matsubara frequencies, fermionic and bosonic ones, respectively. We will also include all $1/(\beta N_k)$ prefactors connected with momentum and frequency sums (with $N_k$ being the number of discrete momenta and $\beta=1/T$) in the definition of the $\sum$ symbol, i.e.
$
\sum_k \equiv \frac{1}{\beta N_k}\sum_{\vec{k},\nu} 
$.
We use the particle-hole notation \cite{RohringerRMP} for the three-frequency, three-momenta objects, i.e.
\begin{align}
&G^{ijlm}_{\sigma_1\ldots\sigma_4}(\tau_1,\tau_2,\tau_3,\tau_4) 
=  \sum_{k,k',q} G_{\sigma_1\ldots\sigma_4}^{kk'q} e^{-i\nu\tau_1} e^{i(\nu+\omega)\tau_2} \nonumber \\ 
&\times e^{-i(\nu'+\omega)\tau_3} e^{i\nu'\tau_4} 
e^{-i\vec{k}\vec{r}_i} e^{i(\vec{k}+\vec{q})\vec{r}_j} e^{-i(\vec{k'}+\vec{q})\vec{r}_l} e^{i\vec{k'}\vec{r}_m},
\end{align}
with $\vec{r}_i$ denoting the position vector of site $i$.

We also make use of the $SU(2)$ symmetry and adopt the following spin notation: $\uparrow\uparrow\uparrow\uparrow = \downarrow\downarrow\downarrow\downarrow\equiv \uparrow\uparrow$, $\uparrow\uparrow\downarrow\downarrow = \downarrow\downarrow\uparrow\uparrow\equiv\uparrow\downarrow$, $\uparrow\downarrow\downarrow\uparrow = \downarrow\uparrow\uparrow\downarrow \equiv \overline{\uparrow\downarrow} = \uparrow\uparrow - \uparrow\downarrow $ for the two-particle Green's function. To keep track of derivatives of some of the following expressions we explicitly write the imaginary time arguments $\tau_1,\ldots,\tau_4$, which are in the end set to $0^+$.

This yields the following expressions of the correlators in terms of Green's functions:

\begin{enumerate}
	\item  Term containing the double derivative
	\begin{align}
	 &{\cal C}_1(i,j) \nonumber \\
   &= \frac{\partial }{\partial \tau_1} \frac{\partial}{\partial \tau_3} \langle  c_{i\uparrow}(\tau_1) c^{\dagger}_{i\uparrow}(\tau_2) c_{j\uparrow} (\tau_3) c^{\dagger}_{j\uparrow} (\tau_4)\rangle  \nonumber \\
  &= -\sum_{k,k',q}\nu(\nu'+\omega)\;\widehat{{G}}_{\uparrow\uparrow}^{kk'q}e^{i\vec{q}{(\vec{r}_i-\vec{r}_j)}}, 
  \label{eq:C1}
	\end{align}
where $\widehat{{G}}$ is the Fourier transform of the correlator introduced in Eq.~\eqref{eq:gbarwave}, now in particle-hole notation for the given spin combination, see also Appendix~\ref{sec:AppA} for the complete expression.
     \item Terms with one derivative (coming from the commutator with the hopping term) with $\delta$ denoting all possible shifts of the site indices $i$, $j$ to the nearest neighbors 
     \begin{align}
     	&{\cal C}_t(i,j)= {\cal C}_{t1}(i,j) + {\cal C}_{t2}(i,j)\nonumber \\
      & = \sum_{\delta} \left[ \frac{\partial}{\partial \tau_3} \langle  c_{i+\delta\uparrow}(\tau_1) c^{\dagger}_{i\uparrow}(\tau_2) c_{j\uparrow} (\tau_3) c^{\dagger}_{j\uparrow} (\tau_4)\rangle  \nonumber \right. \\
& + \left. \frac{\partial }{\partial \tau_1}  \langle  c_{i\uparrow}(\tau_1) c^{\dagger}_{i\uparrow}(\tau_2) c_{j+\delta\uparrow} (\tau_3) c^{\dagger}_{j\uparrow} (\tau_4)\rangle \right] \nonumber \\
            & = \frac{1}{t}\sum_{k,k',q}i(\nu'+\omega) \;\overline{{G}}_{\uparrow\uparrow}^{kk'q}e^{i\vec{q}{(\vec{r}_i-\vec{r}_j)}}\varepsilon_{\vec{k}}\nonumber \\
                  & + \frac{1}{t}\sum_{k,k',q}i\nu \;\widetilde{{G}}_{\uparrow\uparrow}^{kk'q}e^{i\vec{q}{(\vec{r}_i-\vec{r}_j)}}\varepsilon_{\vec{k}'+\vec{q}} ,
     \end{align}
     where $\varepsilon_{\vec{k}}$ is the free electron dispersion relation (for the square lattice $\varepsilon_{\vec{k}}=-2t(\cos k_x + \cos k_y)$) and the modified Green's functions $\overline{{G}}$ and $\widetilde{{G}}$ are given in the Appendix~\ref{sec:AppA}.    
     \item  Terms with one derivative from the chemical potential term  
   \begin{align}
    & {\cal C}_{\mu} (i,j) = {\cal C}_{\mu1} (i,j) + {\cal C}_{\mu2} (i,j) \nonumber \\
        & = \left. \frac{\partial}{\partial \tau_3} \langle  c_{i\uparrow}(\tau_1) c^{\dagger}_{i\uparrow}(\tau_2) c_{j\uparrow} (\tau_3) c^{\dagger}_{j\uparrow} (\tau_4)\rangle \right. \nonumber  \\
    & + \left. \frac{\partial}{\partial \tau_1} \langle  c_{i\uparrow}(\tau_1) c^{\dagger}_{i\uparrow}(\tau_2) c_{j\uparrow} (\tau_3) c^{\dagger}_{j\uparrow} (\tau_4)\rangle \right.\nonumber \\
       & = -\sum_{k,k',q}i(\nu'+\omega) \;\overline{{G}}_{\uparrow\uparrow}^{kk'q}e^{i\vec{q}{(\vec{r}_i-\vec{r}_j)}}\nonumber\\
                  & - \sum_{k,k',q}i\nu \;\widetilde{{G}}_{\uparrow\uparrow}^{kk'q}e^{i\vec{q}{(\vec{r}_i-\vec{r}_j)}}.  
   \end{align}  
     \item One more term with a derivative 
     \begin{align}
    	{\cal C}_4(i,j) & = \frac{\partial}{\partial \tau_1}  \langle  c_{j \downarrow} (\tau_1)c^\dagger_{i\downarrow} (\tau_2)c_{i \uparrow} (\tau_3)c^{\dagger}_{i \uparrow}(\tau_4)\rangle  \nonumber \\
      & = - \sum_{k,k',q}i\nu \;\widetilde{{G}}_{\uparrow\downarrow}^{kk'q}e^{i\vec{k}{(\vec{r}_i-\vec{r}_j)}}. 
     \end{align}
  
     \item The double shift term ($\Delta$ also denotes all possible shifts to nearest neighbors). Since the following terms do not contain time derivatives, we omit the imaginary-time arguments.  
     \begin{align} 
     {\cal C}_{t^2}(i,j)  =&  \sum_{\delta,\Delta}  \langle  c_{i+\delta\uparrow} c^{\dagger}_{i\uparrow} c_{j+\Delta\uparrow}  c^{\dagger}_{j\uparrow} \rangle  \nonumber \\
     =& \frac{1}{t^2}\sum_{k,k',q} \;{{G}}_{\uparrow\uparrow}^{kk'q}e^{i\vec{q}{(\vec{r}_i-\vec{r}_j)}}\varepsilon_{\vec{k}}\;\varepsilon_{\vec{k}'+\vec{q}}. 
     \end{align}
     \item Single shift terms 
      \begin{align}
       {\cal C}_{\mu t}(i,j)& =    {\cal C}_{\mu t1}(i,j) +    {\cal C}_{\mu t2}(i,j) \nonumber\\&=\sum_{\delta} \left[   \langle  c_{i+\delta\uparrow} c^{\dagger}_{i\uparrow} c_{j\uparrow}  c^{\dagger}_{j\uparrow} \rangle +  \langle  c_{i\uparrow} c^{\dagger}_{i\uparrow} c_{j+\delta\uparrow} c^{\dagger}_{j\uparrow} \rangle \right] \nonumber\\
       &= - \frac{1}{t}\sum_{k,k',q} \;{{G}}_{\uparrow\uparrow}^{kk'q}e^{i\vec{q}{(\vec{r}_i-\vec{r}_j)}}\varepsilon_{\vec{k}} \nonumber\\ 
   &  - \frac{1}{t}\sum_{k,k',q} \;{{G}}_{\uparrow\uparrow}^{kk'q}e^{i\vec{q}{(\vec{r}_i-\vec{r}_j)}}\varepsilon_{\vec{k}'+\vec{q}}.
      \end{align}
      \begin{align}
            {\cal C}_{12}(i,j)& =  \sum_{\delta} \langle   c_{j+\delta\downarrow} c^{\dagger}_{i\downarrow} c_{i\uparrow} c^{\dagger}_{i\uparrow} \rangle \nonumber\\
       &= - \frac{1}{t}\sum_{k,k',q} \;{{G}}_{\uparrow\downarrow}^{kk'q}e^{i\vec{k}{(\vec{r}_i-\vec{r}_j)}}\varepsilon_{\vec{k}}. 
      \end{align}
\item 4-leg hopping terms
            \begin{align}
     	{\cal C}_{8A}(i,j)&=\langle  c_{j\downarrow} c^{\dagger}_{i\downarrow}  c_{i\uparrow} c^{\dagger}_{i\uparrow}   \rangle =\sum_{k,k',q} \;{{G}}_{\uparrow\downarrow}^{kk'q}e^{i\vec{k}{(\vec{r}_i-\vec{r}_j)}}
     \end{align}
      \begin{align}
     	{\cal C}_{8B}(i,j)&=\langle  c_{j\downarrow} c^{\dagger}_{i\downarrow}  c_{j\uparrow} c^{\dagger}_{j\uparrow}   \rangle \nonumber\\ 
      & =\sum_{k,k',q} \;{{G}}_{\uparrow\downarrow}^{kk'q}e^{i(\vec{k}+\vec{q}){(\vec{r}_i-\vec{r}_j)}} \; .
     \end{align}
     \item  The following terms are directly representable with the help of susceptibilities $\chi^q_\alpha$ (magnetic for $\alpha = m$, density for $\alpha = d$ and pairing $\alpha = s$, see Appendix~\ref{sec:AppB} for the definitions) and average occupation per site $\langle n_i \rangle$ (summed over spin, i.e. $\langle n_i \rangle = \langle n_{i\uparrow} +  n_{i\downarrow} \rangle$ )
     \begin{align}
     {\cal C}_{5}(i,j)&=\langle 	c_{i\uparrow} c^{\dagger}_{i\uparrow}  c_{j\uparrow} c^{\dagger}_{j\uparrow} \rangle = \sum_{k,k',q} \;{{G}}_{\uparrow\uparrow}^{kk'q}e^{i\vec{q}{(\vec{r}_i-\vec{r}_j)}} \nonumber\\ & = (1- \langle n_{i\uparrow} \rangle)^2 + \frac{1}{4}\sum_q (\chi_d^q+\chi_m^q) e^{i\vec{q}{(\vec{r}_i-\vec{r}_j)}}
     \end{align}
     \begin{align}
      {\cal C}_{6}(i,j)&=\langle 	c_{i\uparrow}  c^{\dagger}_{i\uparrow} c_{j\downarrow} c^{\dagger}_{j\downarrow} \rangle =\sum_{k,k',q} \;{{G}}_{\uparrow\downarrow}^{kk'q}e^{i\vec{q}{(\vec{r}_i-\vec{r}_j)}} \nonumber\\ 
      & =(1- \langle n_{i\uparrow} \rangle)^2  + \frac{1}{4}\sum_q (\chi_d^q-\chi_m^q) e^{i\vec{q}{(\vec{r}_i-\vec{r}_j)}}
     \end{align}
      \begin{align}
     {\cal C}_{7}(i,j) &=\langle 	c_{i\uparrow} c^{\dagger}_{i\uparrow} c_{i\downarrow} c^{\dagger}_{i\downarrow} \rangle  =\sum_{k,k',q} \;{{G}}_{\uparrow\downarrow}^{kk'q}\nonumber\\ 
      &=(1- \langle n_{i\uparrow} \rangle)^2  + \frac{1}{4}\sum_q (\chi_d^q-\chi_m^q) 
     \end{align}
 \begin{align}
       	& {\cal C}_{10}(i,j)=-\langle c_{j\downarrow} c^{\dagger}_{i\uparrow}   c_{j\uparrow}  c^{\dagger}_{i\downarrow}\rangle \nonumber\\ 
      &=-\sum_{k,k',q} \;{{G}}_{\overline{\uparrow\downarrow}}^{kk'q}e^{i(\vec{k}+\vec{k}'+\vec{q}){(\vec{r}_i-\vec{r}_j)}} = \sum_q \chi_s^q \;e^{i\vec{q}{(\vec{r}_i-\vec{r}_j)}}
     \end{align}
      \begin{align}
        	&{\cal C}_{11}(i,j)=\langle   c_{i\uparrow} c^{\dagger}_{i\downarrow} c_{j\downarrow} c^{\dagger}_{j\uparrow}\rangle= \sum_{k,k',q} \;{{G}}_{\overline{\uparrow\downarrow}}^{kk'q}e^{i\vec{q}  (\vec{r}_i-\vec{r}_j)}\nonumber\\ 
      &=\frac{1}{2}\sum_q \chi_m^q e^{i\vec{q}{(\vec{r}_i-\vec{r}_j)}}
         \end{align}
     \item The last two terms needed are simply related to the one-particle Green's function (since $G_{\uparrow\uparrow}^k = G_{\downarrow\downarrow}^k$, we omit the spin index)
  \begin{align}
     	{\cal C}_{9}(i,j)=\langle c_{j \uparrow} c^{\dagger}_{i\uparrow}  \rangle=-\sum_k G^k e^{i\vec{k}{(\vec{r}_i-\vec{r}_j)}},
    \end{align}
      \begin{align}
     	{\cal C}_{13}(i,j)=\langle c_{i\uparrow} c^\dagger_{i\uparrow}\rangle =-\sum_k G^k = 1 - \langle n_{i\uparrow}\rangle.
      \label{eq:C13}
    \end{align}
\end{enumerate} 
The expressions for the correlators ${\cal C}_{4}$, ${\cal C}_{8A}$, ${\cal C}_{8B}$, ${\cal C}_{12}$, ${\cal C}_{\mu}$, and  ${\cal C}_{\mu t}$ can be reformulated in terms of three-point correlation functions (see Appendix~\ref{sec:AppB}). Only  ${\cal C}_{1}$,  ${\cal C}_{t}$, and  ${\cal C}_{t^2}$ need the knowledge of the full two-particle Green's function
or the two-particle vertex.

From the one-and two-particle Green's functions, we can thus  first calculate the correlators through Eqs.~(\ref{eq:C1}-~\ref{eq:C13}) for two arbitrary sites $i$ and $j$. These in turn yield via Eqs.~(\ref{eq:rho116}-\ref{eq:rho1212}) [where the site indices $i$ and $j$ have been dropped]
all elements of the two-site reduced density matrix. Knowing the relation between two-site density matrix elements and Green's functions allows us to compute the density matrix from Green's function methods, in particular those that also compute the two-particle vertex.


\section{Results}
\label{sec:results}

As an example of application and validation of our equations, we show results for the mutual information $I$ and entanglement negativity $N$ between two distinct sites $i$ and $j$ for the   Hubbard model~\eqref{eq:Hubbard}. We consider the following geometries: two sites only, four sites arranged in a $2\times 2$ cluster (here we can choose the two sites either as nearest neighbors or as diagonal next-nearest neighbors), and six sites arranged in a ring (here we have nearest, second nearest and third nearest neighbors). For the $2\times 2$ cluster we use periodic boundary conditions applied both in $x$ and $y$ direction -- it is  then equivalent to a $4$-site ring with the hopping equal to $2t$. The results were obtained from exact diagonalization (the overall dimension of the Fock space for the $6$-site ring is $2^{12}=4096$, so the full numerical diagonalization is easily reachable). For the $2\times 2$ cluster at $T>0$ Eqs.~\eqref{eq:C1}-\eqref{eq:C13} were evaluated both directly through expectation values of the operators in real space (calculating the derivatives with the use of the commutator), and through the momentum and frequency sums of the Green's functions obtained also with exact diagonalization from the \texttt{Fermions.jl} package written in \texttt{Julia}~\cite{bezanson2017julia} and introduced in Ref.~\onlinecite{Wallerberger2022}. Since the relations of $\cal{C}$ correlators to Green's functions are exact, the numerical computations only confirmed the feasibility of performing the momentum and frequency sums needed in the representations through Green's function (for examples of convergence with respect to the number of Matsubara frequencies, see Appendix~\ref{App:convergence}) and provided for a double check of the equations presented in this paper. As  an additional check, for all clusters considered the elements of the two-site reduced density matrix  were also evaluated directly from Eqs.~\eqref{eq:diag}-\eqref{eq:non-diag} and compared to the results obtained from Eqs.~\eqref{eq:C1}-\eqref{eq:C13}. Eqs.~\eqref{eq:diag}-\eqref{eq:non-diag} were also used to compute the results at $T=0$. The imaginary time/frequency formalism was applied only to $T>0$ cases.  

\begin{figure}[tb]
\includegraphics[width=\linewidth]{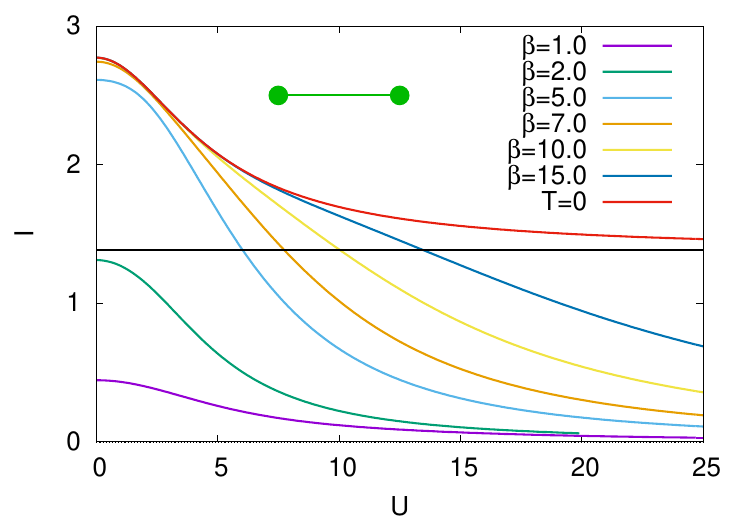}
\caption{Mutual information in the two-site Hubbard model at half filling ($\mu=U/2$) between the two sites as a function of the interaction $U$ for different values of the inverse temperature $\beta=1/T$. The horizontal solid black line denotes $I=2 \ln 2$. Here and in the following figures, all energies and temperatures are in units of $t$.
}
\label{fig:MI}
\end{figure}

\begin{figure}[tb]
\includegraphics[width=\linewidth]{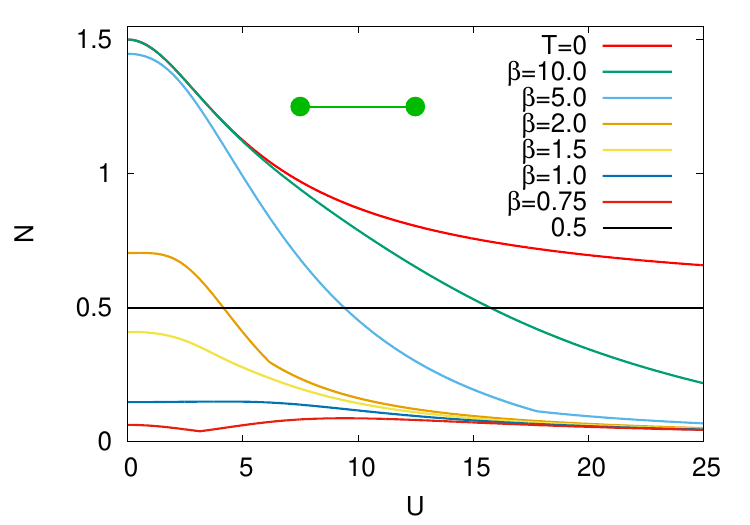}
\caption{Entanglement negativity of the two-site Hubbard model at half filling ($\mu=U/2$) between the two sites as a function of the interaction $U$ for different values of the inverse temperature $\beta=1/T$. The horizontal black line denotes the $T=0$ strong coupling limit.}
\label{fig:N}
\end{figure}

Let us start the discussion of the results with the two-site model
for which the mutual information $I$ is plotted in Fig.~\ref{fig:MI}
and the negativity in  Fig.~\ref{fig:N}. Here and in the following, all parameters given in the plots ($U$, $\mu$, temperature $T$) are in units of the hopping amplitude, i.e. $t\equiv 1$ and $k_B\equiv1$.
In this special two-site case, the calculated two-site density matrix corresponds to the full density matrix of the system, and $I$ is twice the von Neumann entanglement entropy at $T=0$ since the full density matrix $S_{A\cup B}=0$ for a pure state. 
For $U=0$, $T=0$, all states on site $A$ are equally populated; thus we have a diagonal density matrix $(S_A)_{mn}=1/4 \; \delta_{mn}$ and  $I=2S_A=2\ln4$. 
For $U\rightarrow \infty$, $T=0$, we have the same but only for the two-spin states and thus $I=2S_A=2\ln2$. Both correspond to maximal entanglement of the electronic and spin states, respectively. For the spin state at $U\rightarrow \infty$ it is the usual singlet state between the two sites; for $U=0$ we simply have the Slater determinant occupied by spin up and down for the bonding state which is maximally entangled.
Increasing $T$ (decreasing $\beta$) suppresses this entanglement in Fig.~\ref{fig:MI} as all possible states become thermally occupied.


\begin{figure}[tb]
\includegraphics[width=\linewidth]{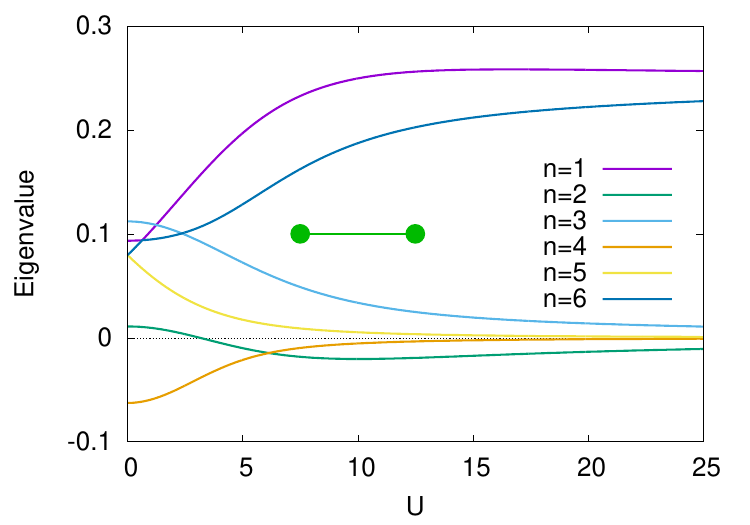}
\caption{Distinct eigenvalues $\epsilon_n$ of the partial transposed density matrix  $\rho^{T_A}$ in the two-site Hubbard model at half filling ($\mu=U/2$) as a function of the interaction $U$ for the inverse temperature $\beta=1/T=0.75$.}
\label{fig:EV}
\end{figure}

The $T=0$ negativity in Fig.~\ref{fig:N} shows a similar behavior as the mutual information in Fig.~\ref{fig:MI}. Notable is a non-monotonous behavior at finite temperatures and even some kinks at intermediate $U$ values. This emerges from eigenvalues of the partial transpose crossing zero, see Fig.~\ref{fig:EV}.

\begin{figure}[tb]
\includegraphics[width=\linewidth]{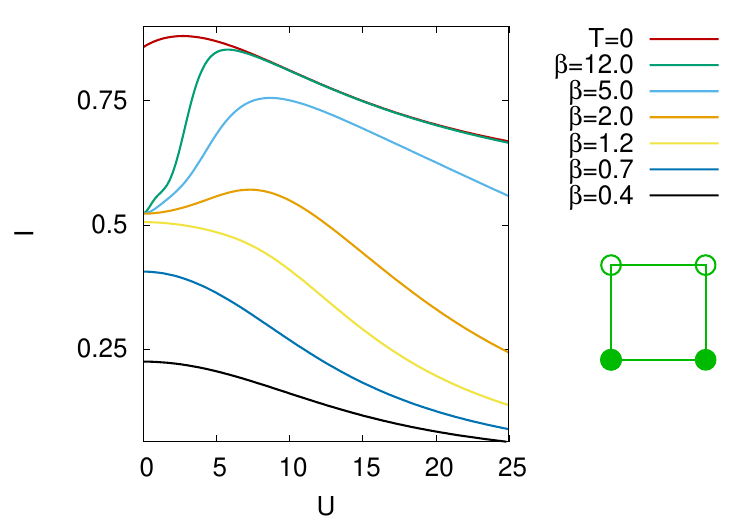}
\caption{Mutual information in the $2\times 2$ Hubbard model at half filling ($\mu = U/2$) between the two neighboring sites (see inset) as a function of the interaction strength $U$ for different inverse temperatures $\beta=1/T$. }
\label{fig:MINN}
\end{figure}

\begin{figure}[tb]
\includegraphics[width=\linewidth]{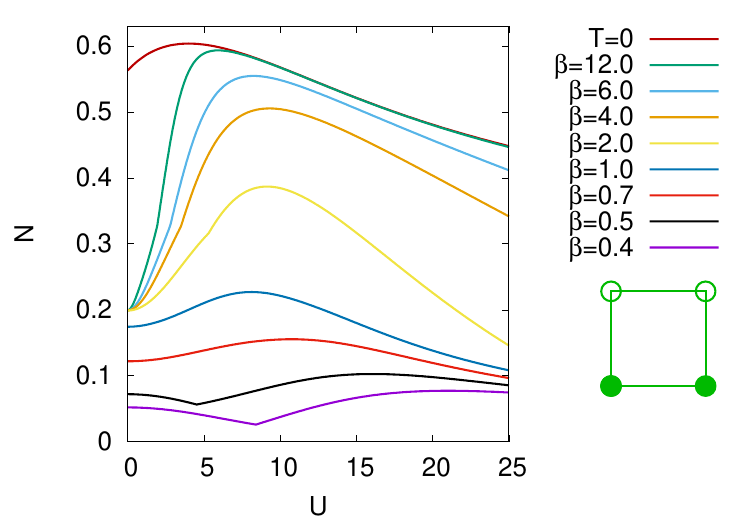}
\caption{Negativity in the $2\times 2$ Hubbard model at half filling ($\mu = U/2$) between the two neighboring sites (see inset) as a function of the interaction strength $U$ for different inverse temperatures $\beta=1/T$. }
\label{fig:NNN}
\end{figure}

\begin{figure}[tb]
\includegraphics[width=\linewidth]{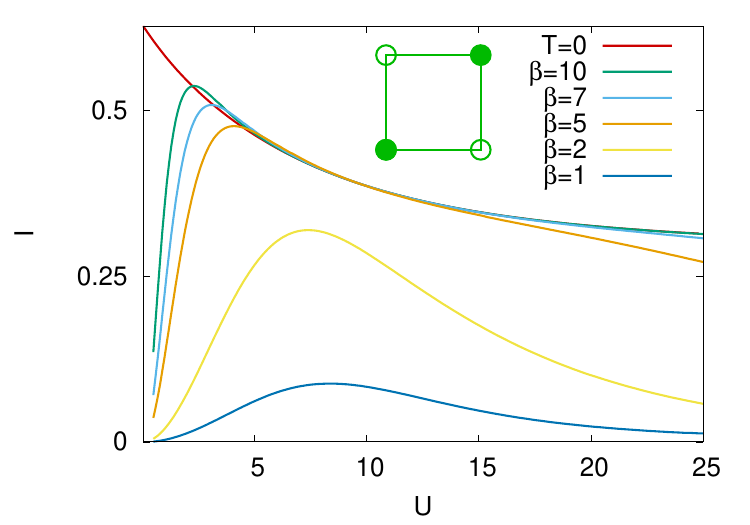}
\caption{Mutual information in the $2\times 2$ Hubbard model at half filling ($\mu = U/2$) between the two corners (see inset) as a function of the interaction strength $U$ for different inverse temperatures $\beta=1/T$.  }
\label{fig:MIdiag}
\end{figure}

\begin{figure}
\includegraphics[width=\linewidth]{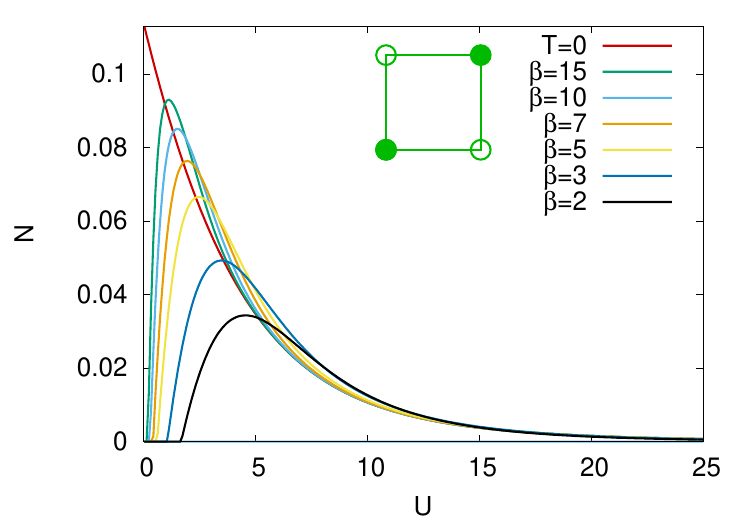}
\caption{Negativity in the $2\times 2$ Hubbard model at half filling ($\mu = U/2$) between the two corners (see inset)  as a function of the interaction strength $U$ for different inverse temperatures $\beta=1/T$.   
\label{fig:Ndiag}}
\end{figure}

For the  $2\times 2$ Hubbard cluster we have two inequivalent arrangements of the two sites considered in the reduced density matrix: neighboring ones and diagonal ones, see insets of Figs.~\ref{fig:MINN}, \ref{fig:NNN} and Figs.~\ref{fig:MIdiag}, \ref{fig:Ndiag}, respectively. Both can be investigated separately.  Let us start discussing the nearest neighbor case.
Here, both the mutual information $I$ in Fig.~\ref{fig:MINN} and the negativity $N$ in Fig.~\ref{fig:NNN} are suppressed compared to the two-site case in Fig.~\ref{fig:MI} and \ref{fig:N}. This is to be expected since the entanglement of neighboring sites competes with entanglement with the other sites, it cannot be perfect anymore for two sites.
Qualitatively different is that  $I$ and $N$ indicate an enhanced entanglement at intermediate $U$ where we have a particularly strong antiferromagnetic coupling.

Even  more different to the nearest neighbor case is $I$ and,  particularly, $N$ for the diagonal neighbors of the $2\times 2$ cluster shown in  Fig.~\ref{fig:MIdiag} and \ref{fig:Ndiag}, respectively. Here $I$ approaches zero for $U\rightarrow 0$; the negativity $N$ is even zero for small but finite $U$'s. 
This can be understood by the vanishing one particle Green's function between every second site in the one-dimensional Hubbard model at $U=0$ \cite{Komnik2017}.
The negativity is also strongly suppressed,  actually goes to zero in the limit $U\rightarrow \infty$,   where the Hubbard model maps onto a spin model.
In contrast, the mutual information is merely reduced.

\begin{figure}[tb]
\includegraphics[width=\linewidth]{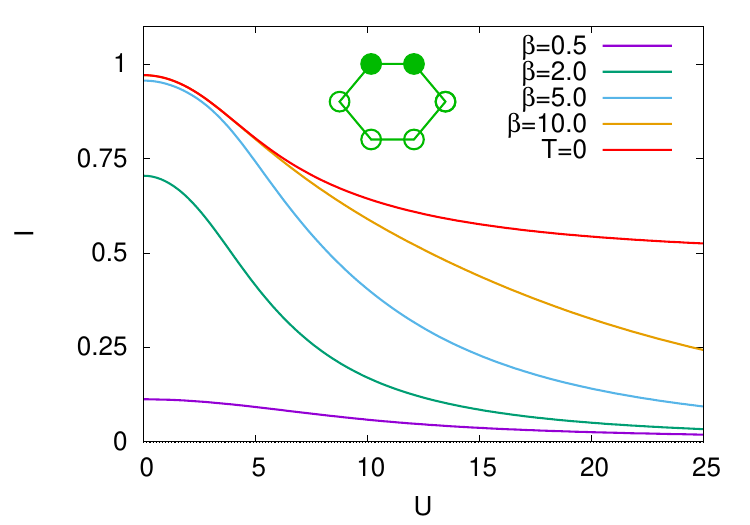}
\caption{Mutual information in the $6$-site Hubbard ring at half filling ($\mu = U/2$) and $t=1$ between the two points (see inset) as a function of the interaction strength $U$ for different inverse temperatures $\beta=1/T$.  }
\label{fig:MI_6_NN}
\end{figure}

\begin{figure}[tb]
\includegraphics[width=\linewidth]{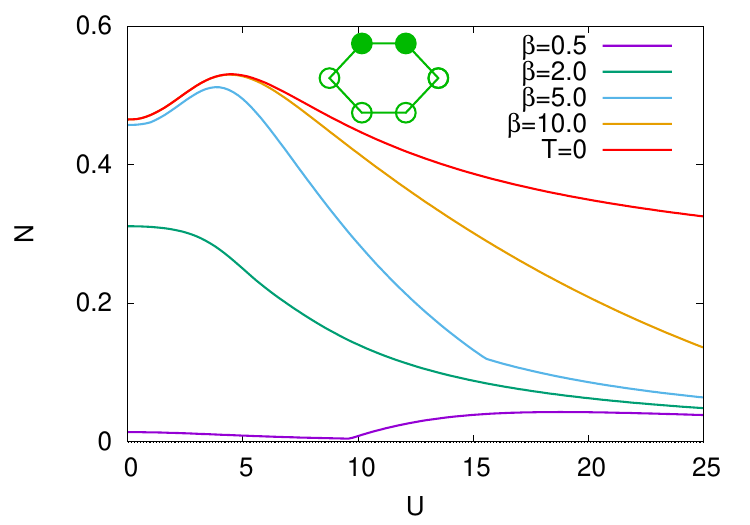}
\caption{ Negativity in the $6$-site Hubbard ring at half filling ($\mu = U/2$) and $t=1$ between the two points (see inset) as a function of the interaction strength $U$ for different inverse temperatures $\beta=1/T$.  }
\label{fig:N_6_NN}
\end{figure}

\begin{figure}[tb]
\includegraphics[width=\linewidth]{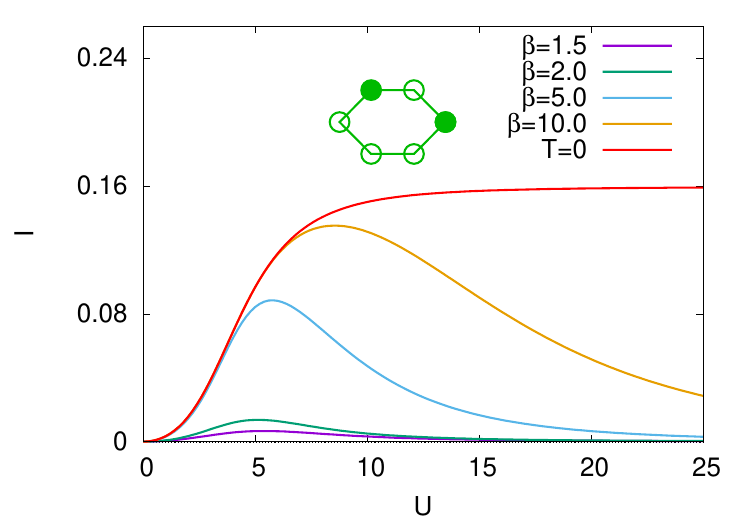}
\caption{Mutual information in the $6$-site Hubbard ring at half filling ($\mu = U/2$) and $t=1$ between the two points (see inset) as a function of the interaction strength $U$ for different inverse temperatures $\beta=1/T$.  }
\label{fig:MI_6_NNN}
\end{figure}

\begin{figure}[tb]
\includegraphics[width=\linewidth]{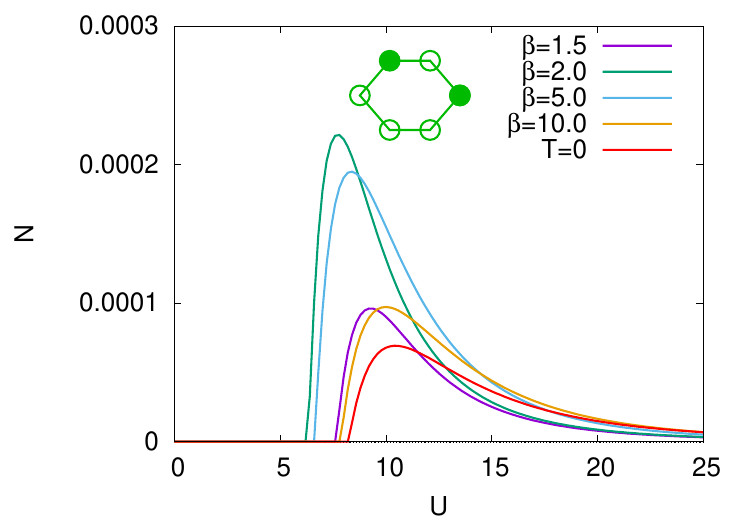}
\caption{ Negativity in the $6$-site Hubbard ring at half filling ($\mu = U/2$) and $t=1$ between the two points (see inset) as a function of the interaction strength $U$ for different inverse temperatures $\beta=1/T$.  }
\label{fig:N_6_NNN}
\end{figure}

\begin{figure}[tb]
\includegraphics[width=\linewidth]{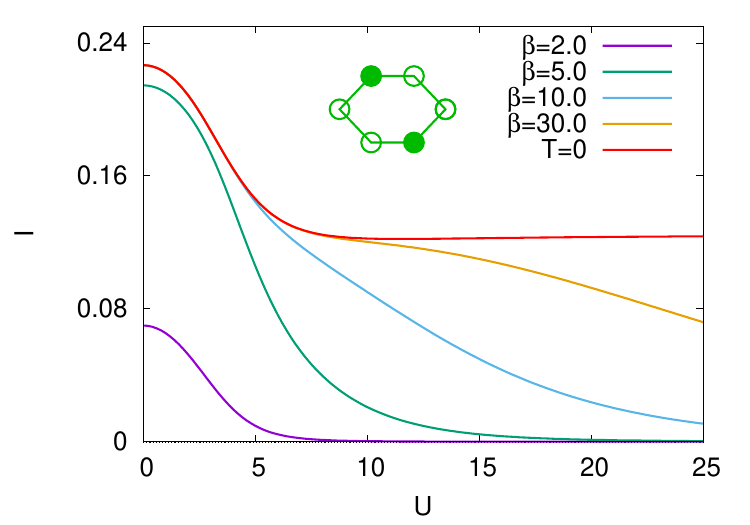}
\caption{Mutual information in the $6$-site Hubbard ring at half filling ($\mu = U/2$) and $t=1$ between the two points (see inset) as a function of the interaction strength $U$ for different inverse temperatures $\beta=1/T$.  }
\label{fig:MI_6_NNNN}
\end{figure}

\begin{figure}[tb]
\includegraphics[width=\linewidth]{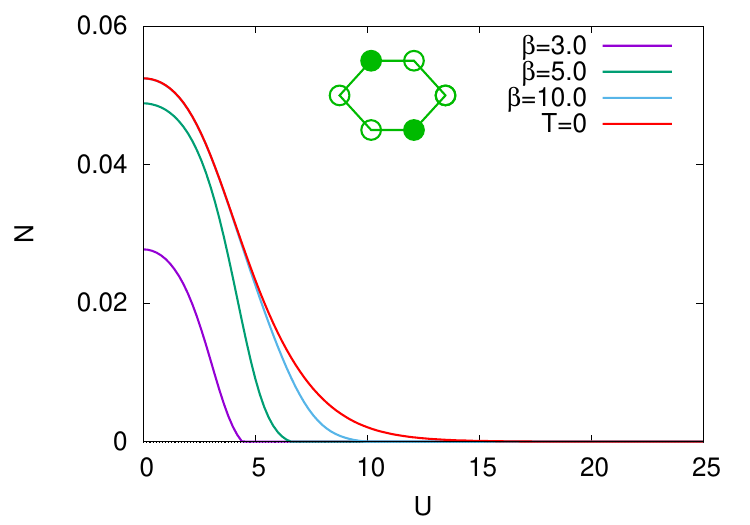}
\caption{ Negativity in the $6$-site Hubbard ring at half filling ($\mu = U/2$) and $t=1$ between the two points (see inset) as a function of the interaction strength $U$ for different inverse temperatures $\beta=1/T$.  }
\label{fig:N_6_NNNN}
\end{figure}

The method can also be applied to more distant sites. To illustrate this point with a simple toy model we introduce the $6$-site ring as a final example. The nearest neighbor mutual information and negativity can be observed in Fig.~\ref{fig:MI_6_NN} and \ref{fig:N_6_NN} respectively. The negativity behaves very similar to the nearest neighbor case of the $2\times 2$ lattice. Only the $T\rightarrow0, U\rightarrow0$ limit is different. While the tight-binding model has a spin degeneracy in the ground state of the four-site model, no such degeneracy exits for the six-site ring as we have three fully filled k-states with no degeneracy. Therefore, the mutual information behaves more closely to that of the dimer at small $U$. The case of entanglement with one intermediate site 
in-between the two sites considered
is displayed in Fig.~\ref{fig:MI_6_NNN} and \ref{fig:N_6_NNN} and has close resemblance to the diagonal entanglement in the $2\times 2$ system, with the same difference for $T\rightarrow0, U\rightarrow0$. We also note that the negativity requires comparably large values of $U$ for the six site ring to detect entanglement. This can be explained by the energy gap that separates the ground state from the first excited state. Therefore also higher temperatures are more entangled than the ground state $T=0$ before even higher temperatures cancel all correlations and entanglement. The farthest distance of $2$ intermediate sites is provided in Fig.~\ref{fig:MI_6_NNNN} and \ref{fig:N_6_NNNN}. For small $U$ and finite $T$ this behaves just like the two-site case, as it has the same spatial symmetry, effectively the intermediate sites lead to a reduced hopping parameter $t$. However this tendency is suppressed at large $U$ where the hopping is suppressed and the behavior can best be compared to the diagonal case in the $2\times 2$ system. Due to the symmetries one would expect this similarity to hold for small $U$ as well. However, here the vanishing one particle Green's function for an odd number of intermediate sites at $U=0$ \cite{Komnik2017} forces a down-turn in the $2\times 2$ system.


\section{Conclusion and outlook}
\label{sec:conclusions}
We have outlined in detail a method for calculating the two-site reduced density matrix from one- and two-particle Green's functions. To this end, we have shown, and also confirmed numerically, that also the expectation values of eight and six fermionic operators in the two-site  density matrix can be evaluated with the help of equations of motion using only four-point correlators.  To connect these to Green's functions, we reformulated the resulting derivatives of the two-particle time-ordered correlators in terms of one- and two-particle Green's functions summed over frequencies. The final expressions can thus be readily used within any Green's function method which calculates or approximates the one- and two-particle Green's function (or the two-particle vertex). Our formalism thus allows us to use two-particle diagrammatic approaches such as the dynamical vertex approximation~\cite{Toschi2007,RohringerRMP}, parquet equations~\cite{DeDominicis,Eckhardt2020}, and the functional RG~\cite{Metzner2012,Hille2020}, and to calculate --in a post processing step--  also entanglement 
measures such as the mutual information and negativity.

The presented derivations were obtained for the local interaction only and nearest-neighbor hopping in the Hubbard model. It is however straightforward to extend it to other Hamiltonians. (There will then appear more terms in the equation of motion originating from the commutator $[H,c_{i\sigma}]$, but otherwise analogous steps can be taken.)   Many of the reduced density matrix elements require only the knowledge of the one-particle Green's function and charge and magnetic susceptibilities. These can be even obtained experimentally. 
It is an open question how important for the entanglement measures are density matrix elements that require the knowledge of the full two-particle vertex and cannot be obtained from susceptibilities or 3-point electron-boson vertices.  To investigate this question, as well as how different approximations  reflect in the entanglement measures, is an interesting direction for future work with the Green's function representation of the matrix elements.
Our work represents a starting point and the results for the two-, four-, and six-site clusters are only confirming the feasibility of the computations and serving as a benchmark. For extended systems the correlators cannot be obtained by direct exact diagonalization, so the Green's function approach offers a possible alternative. 

In a broader context, having the possibility to compute the reduced density matrix as a mere postprocessing step of a Green's function method paves the way for a much more widespread calculation of  entanglement in strongly correlated fermion systems.
We hope that this will eventually improve our understanding of entanglement in fermionic systems as well as of the differences between and appropriateness of various entanglement measures \cite{Ryu2017,Ryu2019,Ding2021,Ding2022,Ding2023,Ding2024} for fermions.

{\it Note added.} Upon completion of the present manuscript we became aware of a related recent paper \cite{Adriano2024}, where the authors apply cluster dynamical mean-field theory
 to calculate the two-point entanglement and correlation measures in the two-dimensional Hubbard model. We think that our methods are complementary to each other, while the method used in \cite{Adriano2024}  gives almost exact results within the block size, our Green-function-based method is better suited to investigate correlations between distant sites.


\begin{acknowledgements}
We thank Christian Mayrhofer for careful reading of the technical part of the manuscript. GR has been supported by  the National Research,
Development and Innovation Office NKFIH under Grant
No. K128989,  No. K146736, by the Quantum Information National Laboratory of Hungary, and by the Hungarian Lóránt Eötvös mobility scholarship;  AK by project V 1018 of the Austrian Science Fund
(FWF); FB and KH by  the
SFB Q-M\&S (FWF project ID F86).
\end{acknowledgements}

\appendix

\section{Time derivative of the Green's function and its representation in Matsubara frequencies}
\label{sec:AppA}
\subsection{Calculating time-ordered derivatives and taking the limit of equal times}

To calculate the 8-point or 6-point equal time correlation functions we need derivatives of already time-ordered correlators, e.g. 
 \begin{equation}
 \frac{\partial }{\partial \tau_3}  \langle  c_1(\tau_1) c^{\dagger}_2(\tau_2) c_3 (\tau_3) c^{\dagger}_4 (\tau_4)\rangle,
 \end{equation}
with $\tau_1\ge\tau_2\ge\tau_3\ge\tau_4$. We can calculate these derivatives directly in $\tau$ or connect them to the derivative of the two-particle Green's function $G_{1234}$. The Green's function includes however the time ordering operator $T$ that cannot be neglected when taking the derivative. 
 To begin with a simpler example, let us first look once again at the derivative of the one-particle Green's function
  \begin{align}
 &\frac{\partial }{\partial \tau_3}  G_{34}(\tau_3)  = -\frac{\partial }{\partial \tau_3} \langle T[c_3(\tau_3) c^{\dagger}_4(\tau_4) \rangle \nonumber \\
 & = -\delta(\tau_3 - \tau_4) \langle c_3(\tau_3) c^{\dagger}_4(\tau_4) \rangle - \delta(\tau_4 - \tau_3) \langle c^{\dagger}_4(\tau_4) c_3(\tau_3)  \rangle \nonumber \\
 & - \langle T [\frac{\partial }{\partial \tau_3}  c_3 (\tau_3) c^{\dagger}_4(\tau_4)] \rangle.
  \end{align}
The terms with the Dirac delta can be collected into the anticommutator (because the times become equal under the delta) and we obtain
  \begin{align}
 &\frac{\partial }{\partial \tau_3}  G_{34}(\tau_3)  = -\delta(\tau_3 - \tau_4)\delta_{34} - \langle T [\frac{\partial }{\partial \tau_3}  c_3 (\tau_3) c^{\dagger}_4(\tau_4)] \rangle.
 \label{eq:gf_derivative}
  \end{align}
For the two-particle Green's function we get $4!=24$ terms, but only some of them contribute to the equal time limit. 
 \begin{align}
 &\frac{\partial }{\partial \tau_3} G_{1234}(\tau_1,\tau_2,\tau_3, \tau_4)  = \frac{\partial }{\partial \tau_3} \langle T [  c_1(\tau_1) c^{\dagger}_2(\tau_2)  c_3 (\tau_3) c^{\dagger}_4 (\tau_4)]\rangle \nonumber \\
&=  \langle T [  c_1(\tau_1) c^{\dagger}_2(\tau_2) \frac{\partial }{\partial \tau_3}  c_3 (\tau_3) c^{\dagger}_4 (\tau_4)]\rangle \nonumber \\
 &- \theta(\tau_1 - \tau_2){\delta(\tau_2-\tau_3)}\theta( \tau_3 - \tau_4) \langle    c_1(\tau_1) c^{\dagger}_2(\tau_2) c_3 (\tau_3) c^{\dagger}_4 (\tau_4)\rangle\nonumber \\
  &- \theta(\tau_1 - \tau_2){\delta(\tau_3-\tau_2)}\theta( \tau_3 - \tau_4) \langle    c_1(\tau_1) c_3 (\tau_3) c^{\dagger}_2(\tau_2)  c^{\dagger}_4 (\tau_4)\rangle\nonumber \\
   &- \theta(\tau_1 - \tau_2)\theta(\tau_2-\tau_3){\delta( \tau_3 - \tau_4)} \langle    c_1(\tau_1) c^{\dagger}_2(\tau_2) c_3 (\tau_3) c^{\dagger}_4 (\tau_4)\rangle\nonumber \\
     &- \theta(\tau_1 - \tau_2)\theta(\tau_2-\tau_3){\delta( \tau_4 - \tau_3)} \langle    c_1(\tau_1) c^{\dagger}_2(\tau_2)  c^{\dagger}_4 (\tau_4)c_3 (\tau_3)\rangle\nonumber \label{eq:tder2PGF} \\
 &+ \textrm{other terms}.
 \end{align} 
In the end we are interested in the limit 
 $\tau_1\to\tau_2\to\tau_3\to\tau_4=0$, so we cannot neglect the four divergent terms above, containing $\delta(\tau_2 - \tau_3)$ or $\delta(\tau_3 - \tau_4)$ (there are 4 more terms with $\delta(\tau_2 - \tau_3)$ or $\delta(\tau_3 - \tau_4)$ that contribute and many other terms that do not contribute in the limit we need). When taking the limit however the divergent terms fall apart into disconnected terms (i.e. products of one-particle Green's function and the Dirac delta). To see this, let us look at the last two lines of Eq.~\eqref{eq:tder2PGF}. We collect the two terms using the symmetric property of the delta distribution ($\delta(x) = \delta(-x)$)
 \begin{align}
    &- \theta(\tau_1 - \tau_2)\theta(\tau_2-\tau_3){\delta( \tau_3 - \tau_4)} \langle    c_1(\tau_1) c^{\dagger}_2(\tau_2) \nonumber \\ &\times  [c_3 (\tau_3) c^{\dagger}_4 (\tau_4) + c^{\dagger}_4 (\tau_4)c_3 (\tau_3)]\rangle. \label{eq:comm} 
 \end{align} 
 The Dirac-$\delta$ makes the times of the last two operatos equal, which allows us to collect two terms into an anticommutator 
 \begin{align}
& c_3 (\tau_3) c^{\dagger}_4 (\tau_3) + c^{\dagger}_4 (\tau_3)c_3 (\tau_3) = e^{H\tau_3}  c_3  c^{\dagger}_4 e^{-H\tau_3}\nonumber \\
& + e^{H\tau_3}  c^{\dagger}_4 c_3  e^{-H\tau_3} = e^{H\tau_3} \left\{ c_3,  c^{\dagger}_4 \right \} e^{-H\tau_3} = \delta_{34}.
 \label{eq:comm2} 
 \end{align} 
 After some rearanging and collecting terms into anticommutators and one-particle Green's functions, all the divergent Dirac-$\delta$ terms of Eq.~(\ref{eq:tder2PGF}) can be rewritten to give 
 \begin{align}
& \lim_{\tau_1\to\tau_2\to\tau_3\to\tau_4=0}\textrm{[divergent {(Dirac-$\delta$) terms of Eq.~(\ref{eq:tder2PGF}) }]} \nonumber \\
&=-\lim_{\tau_1\to0^+} \left [G_{12}(\tau_1)  \right] \lim_{\tau_3\to0^+}\left [ \delta(\tau_3)\delta_{34} \right] \nonumber \\
&+\lim_{\tau_1\to0^+} \left [G_{14}(\tau_1)  \right] \lim_{\tau_3\to0^-}\left [ \delta(\tau_3)\delta_{32}   \right]. 
\end{align}
For later use, we will do one more step and replace the $\delta$-distributions by the derivative of the one-particle Green's function from Eq.~\eqref{eq:gf_derivative} to obtain
\begin{align}
& \lim_{\tau_1\to\tau_2\to\tau_3\to\tau_4=0}\textrm{[divergent {(Dirac-$\delta$) terms of Eq.~(\ref{eq:tder2PGF}) }]} \nonumber \\
&=\lim_{\tau_1\to0^+} \left [G_{12}(\tau_1)  \right] \lim_{\tau_3\to0^+}\left [\frac{\partial }{\partial \tau_3}  G_{34}(\tau_3) + \left \langle\frac{\partial }{\partial \tau_3}  c_3 (\tau_3)c^{\dagger}_4 \right\rangle \right]  \nonumber \\
&-\lim_{\tau_1\to0^+} \left [G_{14}(\tau_1)  \right] \lim_{\tau_3\to0^-}\left [\frac{\partial }{\partial \tau_3}  G_{32}(\tau_3) + \left \langle\frac{\partial }{\partial \tau_3}  c_3 (\tau_3) c^{\dagger}_2 \right\rangle \right].
 \end{align}
 
Using the above, we obtain for the equal-time correlation function that we need the following expression:
\begin{align}
& \lim_{\tau_1\to\tau_2\to\tau_3\to\tau_4=0} \langle    c_1(\tau_1) c^{\dagger}_2(\tau_2) \frac{\partial }{\partial \tau_3}  c_3 (\tau_3) c^{\dagger}_4 (\tau_4)\rangle \nonumber \\
& =  \lim_{\tau_1\to\tau_2\to\tau_3\to\tau_4=0} \left [ \frac{\partial }{\partial \tau_3} G_{1234}(\tau_1,\tau_2,\tau_3, \tau_4) \right]\nonumber \\
& -\lim_{\tau_1\to0^+} \left [G_{12}(\tau_1)  \right] \lim_{\tau_3\to0^+}\left [\frac{\partial }{\partial \tau_3}  G_{34}(\tau_3) + \left \langle\frac{\partial }{\partial \tau_3}  c_3 (\tau_3) c^{\dagger}_4 \right\rangle \right]  \nonumber \\
&+\lim_{\tau_1\to0^+} \left [G_{14}(\tau_1)  \right] \lim_{\tau_3\to0^-}\left [\frac{\partial }{\partial \tau_3}  G_{32}(\tau_3) +\left \langle\frac{\partial }{\partial \tau_3}  c_3 (\tau_3) c^{\dagger}_2 \right\rangle \right] \nonumber \\
& \equiv \lim_{\tau_1\to\tau_2\to\tau_3\to\tau_4=0} \left [ \frac{\partial }{\partial \tau_3} \overline{G}_{1234}(\tau_1,\tau_2,\tau_3, \tau_4) \right],
 \end{align}
 where we introduce a notation  $\overline{G}$ for this modified correlator. Let us look more carefully at the derivative of $G_{1234}$. We can 
divide $G$ into connected and disconnected parts
\begin{equation}
G_{1234} = G_{12}G_{34} - G_{14}G_{32} + G^{conn}_{1234}
\end{equation}
and perform the derivative:
\begin{equation}
\frac{\partial }{\partial \tau_3} G_{1234} = G_{12}\frac{\partial }{\partial \tau_3}G_{34} - G_{14}\frac{\partial }{\partial \tau_3}G_{32} + \frac{\partial }{\partial \tau_3}G^{conn}_{1234}.
\end{equation}
 Under the limit of ${\tau_1\to\tau_2\to\tau_3\to\tau_4=0}$ the modified correlator  $\overline{G}$ can be expressed by the connected part of $G$ plus additional disconnected terms
\begin{align}
&\lim_{\tau_1\to\tau_2\to\tau_3\to\tau_4=0} \left [\frac{\partial }{\partial \tau_3} \overline{G}_{1234} \right ] = \lim_{\tau_1\to\tau_2\to\tau_3\to\tau_4=0} \left [\frac{\partial }{\partial \tau_3} G^{conn}_{1234} \right] \nonumber \\
& -\lim_{\tau_1\to0^+} \left [G_{12}(\tau_1)  \right] \lim_{\tau_3\to0^+}\left [ \left \langle\frac{\partial }{\partial \tau_3}  c_3(\tau_3)  c^{\dagger}_4 \right\rangle \right]  \nonumber \\
& +\lim_{\tau_1\to0^+} \left [G_{14}(\tau_1)  \right] \lim_{\tau_3\to0^-}\left [ \left \langle\frac{\partial }{\partial \tau_3}  c_3 (\tau_3) c^{\dagger}_2 \right\rangle \right], 
 \end{align}
 from which we can define
\begin{align}
\overline{G}_{1234}  = G^{conn}_{1234} - G_{12}\langle c_3  c^{\dagger}_4 \rangle +G_{14}\langle c_3  c^{\dagger}_2 \rangle,
 \end{align} 
 where we skipped the time arguments for simplicity. We will show later, in the frequency notation, that the derivative of this correlator does not have divergent terms in the limit of ${\tau_1\to\tau_2\to\tau_3\to\tau_4=0}$. The derivatives over imaginary time can be expressed as multiplication with Matsubara frequency for the Fourier transformed correlators. In the equal time limit, one needs then to sum the Fourier transformed expressions over all Matsubara frequencies (see later).

We can do the same with the derivative over $\tau_1$ to obtain: 
\begin{align}
& \lim_{\tau_1\to\tau_2\to\tau_3\to\tau_4=0} \langle \frac{\partial }{\partial \tau_1}    c_1(\tau_1) c^{\dagger}_2(\tau_2)  c_3 (\tau_3) c^{\dagger}_4 (\tau_4)\rangle \nonumber \\
& =  \lim_{\tau_1\to\tau_2\to\tau_3\to\tau_4=0} \left [ \frac{\partial }{\partial \tau_1} G_{1234}(\tau_1,\tau_2,\tau_3, \tau_4) \right]\nonumber \\
& -\lim_{\tau_1\to0^+} \left [ \frac{\partial }{\partial \tau_1}G_{12}(\tau_1) + \left \langle\frac{\partial }{\partial \tau_1}  c_1(\tau_31)  c^{\dagger}_2 \right\rangle \right] \lim_{\tau_3\to0^+}\left [ G_{34}(\tau_3) \right]  \nonumber \\
&+\lim_{\tau_1\to0^+} \left [ \frac{\partial }{\partial \tau_1} G_{14}(\tau_1) + \left \langle\frac{\partial }{\partial \tau_1}  c_1(\tau_1)  c^{\dagger}_4 \right\rangle \right] \lim_{\tau_3\to0^-}\left [  G_{32}(\tau_3) \right] \nonumber \\
& \equiv \lim_{\tau_1\to\tau_2\to\tau_3\to\tau_4=0} \left [ \frac{\partial }{\partial \tau_1} \widetilde{G}_{1234}(\tau_1,\tau_2,\tau_3, \tau_4) \right],
 \end{align}
where we introduced the correlator $\widetilde{G}_{1234}$, which can again be expressed with the help of connected part of the two-particle Green's function
\begin{align}
&\lim_{\tau_1\to\tau_2\to\tau_3\to\tau_4=0} \left [\frac{\partial }{\partial \tau_1} \widetilde{G}_{1234} \right ] = \lim_{\tau_1\to\tau_2\to\tau_3\to\tau_4=0} \left [\frac{\partial }{\partial \tau_1} G^{conn}_{1234} \right] \nonumber \\
& -\lim_{\tau_1\to0^+} \left [  \left \langle\frac{\partial }{\partial \tau_1}  c_1(\tau_1)  c^{\dagger}_2 \right\rangle \right] \lim_{\tau_3\to0^+}\left [ G_{34}(\tau_3) \right]  \nonumber \\
&+\lim_{\tau_1\to0^+} \left [  \left \langle\frac{\partial }{\partial \tau_1}  c_1(\tau_1)  c^{\dagger}_4 \right\rangle \right] \lim_{\tau_3\to0^-}\left [  G_{32}(\tau_3) \right]. 
 \end{align}
 We then obtain (skipping again the time arguments)
\begin{align}
\widetilde{G}_{1234}  = G^{conn}_{1234} - \langle c_1  c^{\dagger}_2 \rangle G_{34}+\langle c_1  c^{\dagger}_4 \rangle G_{32}.
 \end{align} 
For the double derivative we need to apply the above reasoning twice, but again we can express the derivative by introducing a modified correlator $\widehat{{G}}_{1234}$
\begin{align}
& \lim_{\tau_1\to\tau_2\to\tau_3\to\tau_4=0} \langle \frac{\partial }{\partial \tau_1}   c_1(\tau_1) c^{\dagger}_2(\tau_2) \frac{\partial }{\partial \tau_3}  c_3 (\tau_3) c^{\dagger}_4 (\tau_4)\rangle \nonumber \\
& \equiv  \lim_{\tau_1\to\tau_2\to\tau_3\to\tau_4=0} \left [ \frac{\partial }{\partial \tau_1}\frac{\partial }{\partial \tau_3} \widehat{{G}}_{1234}(\tau_1,\tau_2,\tau_3, \tau_4) \right]\nonumber \\
&= \lim_{\tau_1\to\tau_2\to\tau_3\to\tau_4=0} \left [\frac{\partial }{\partial \tau_1}\frac{\partial }{\partial \tau_3} G^{conn}_{1234} \right] \nonumber \\
& +\lim_{\tau_1\to0^+} \left [  \left \langle\frac{\partial }{\partial \tau_1}  c_1 (\tau_1) c^{\dagger}_2 \right\rangle \right] \lim_{\tau_3\to0^+}\left [ \left\langle\frac{\partial }{\partial \tau_3}  c_3  (\tau_3)c^{\dagger}_4 \right\rangle \right]  \nonumber \\
&-\lim_{\tau_1\to0^+} \left [  \left \langle\frac{\partial }{\partial \tau_1}  c_1 (\tau_1) c^{\dagger}_4 \right\rangle \right] \lim_{\tau_3\to0^-}\left [ \left\langle\frac{\partial }{\partial \tau_3}  c_3  (\tau_3)c^{\dagger}_2 \right\rangle \right]. 
 \end{align}
 From which we get:
\begin{align}
{\widehat{G}}_{1234}  = G^{conn}_{1234} + \langle c_1  c^{\dagger}_2 \rangle \langle c_3  c^{\dagger}_4 \rangle -\langle c_1  c^{\dagger}_4 \rangle \langle c_3  c^{\dagger}_2 \rangle .
 \end{align} 

Let us stress that 
${\widehat{G}}_{1234}$,  
${\overline{G}}_{1234}$, and
$\widetilde{{G}}_{1234}$ can all be calculated from the one- and two-particle Green's functions.

\subsection{Imaginary time derivatives in Matsubara frequency representation }

Since we later want to use $G_{1234}$ in frequency space, let us explicitly evaluate the $\tau$ derivatives of the Fourier transformed quantities. For the one particle Green's function we obtain:
\begin{equation}
    \frac{\partial }{\partial \tau_1}  G_{12}(\tau_1) = \frac{1}{\beta}\sum_\nu (-i\nu) G_{12}(\nu)e^{-i\nu\tau_1}. 
\end{equation}
In the limit of $\tau_1 \to 0^+$ the Matsubara sum diverges. But we are interested in another derivative
\begin{align}
     &\left \langle\frac{\partial }{\partial \tau_1}  c_1 (\tau_1) c^{\dagger}_2 \right\rangle = 
     -\frac{\partial }{\partial \tau_1}  G_{12}(\tau_1) - \delta(\tau_1)\delta_{12}
     \nonumber \\
     &= \frac{1}{\beta}\sum_\nu i\nu \left [G_{12}(\nu) - \frac{1}{i\nu}\delta_{12} \right ] e^{-i\nu\tau_1}. 
\end{align}
The above Matsubara sum converges in the limit of $\tau_1 \to 0^+$. Following the same procedure, we can express the modified correlators $\overline{G}$, $\widetilde{G}$, and $\widehat{{G}}$ through the following frequency sums:
\begin{align}
&\lim_{\tau_1\to\tau_2\to\tau_3\to\tau_4=0} \left [\frac{\partial }{\partial \tau_3} \overline{G}_{1234} \right ] \nonumber \\
& = \frac{1}{\beta^3}\sum_{\nu\nu'\omega} \left [ -i(\nu'+\omega)G_{1234}^{conn\;\nu\nu'\omega}\right]\nonumber \\
&- \frac{1}{\beta}\sum_\nu G_{12}(\nu) e^{-i\nu0^+}  \frac{1}{\beta}\sum_{\nu'} i\nu' \left [G_{34}(\nu') - \frac{\delta_{34}}{i\nu'} \right ] e^{-i\nu'0^+} \nonumber \\
&+ \frac{1}{\beta}\sum_\nu G_{14}(\nu) e^{-i\nu0^+}  \frac{1}{\beta}\sum_{\nu'} i\nu' \left [G_{32}(\nu') - \frac{\delta_{23}}{i\nu'} \right ] e^{-i\nu'0^-}\nonumber \\
& \equiv \frac{1}{\beta^3} \sum_{\nu\nu'\omega} \left [ -i(\nu'+\omega)\overline{G}_{1234}^{\nu\nu'\omega}\right],
\end{align}
\begin{align}
&\lim_{\tau_1\to\tau_2\to\tau_3\to\tau_4=0} \left [\frac{\partial }{\partial \tau_1} \widetilde{G}_{1234} \right ]= \frac{1}{\beta^3}\sum_{\nu\nu'\omega} \left [ -i\nu G_{1234}^{conn\;\nu\nu'\omega}\right]\nonumber \\
&- \frac{1}{\beta}\sum_\nu i\nu \left [G_{12}(\nu)- \frac{\delta_{12}}{i\nu} \right] e^{-i\nu0^+}  \frac{1}{\beta}\sum_{\nu'} G_{34}(\nu')   e^{-i\nu'0^+} \nonumber \\
&+ \frac{1}{\beta}\sum_\nu  i\nu \left [G_{14}(\nu)- \frac{\delta_{14}}{i\nu} \right ] e^{-i\nu0^+}  \frac{1}{\beta}\sum_{\nu'} G_{32}(\nu')  e^{-i\nu'0^-}\nonumber \\
& \equiv \frac{1}{\beta^3} \sum_{\nu\nu'\omega} \left [ -i\nu\widetilde{G}_{1234}^{\nu\nu'\omega}\right],
\end{align}
\begin{align}
&\lim_{\tau_1\to\tau_2\to\tau_3\to\tau_4=0} \left [\frac{\partial }{\partial \tau_1}\frac{\partial }{\partial \tau_3} \widehat{{G}}_{1234} \right ] \nonumber \\
& = \frac{1}{\beta^3}\sum_{\nu\nu'\omega} \left [ \nu(\nu'+\omega)G_{1234}^{conn\;\nu\nu'\omega}\right]\nonumber \\
&+ \frac{1}{\beta^2}\sum_{\nu\nu'} i\nu \left [G_{12}(\nu)- \frac{\delta_{12}}{i\nu} \right ] e^{-i\nu0^+}  i\nu' \left [G_{34}(\nu') - \frac{\delta_{34}}{i\nu'} \right ] e^{-i\nu'0^+} \nonumber \\
&- \frac{1}{\beta^2}\sum_{\nu\nu'} i\nu \left [G_{14}(\nu)- \frac{\delta_{14}}{i\nu} \right ] e^{-i\nu0^+}  i\nu' \left [G_{32}(\nu') - \frac{\delta_{23}}{i\nu'} \right ] e^{-i\nu'0^-}\nonumber \\
& \equiv \frac{1}{\beta^3} \sum_{\nu\nu'\omega} \left [ \nu(\nu'+\omega)\widehat{{G}}_{1234}^{\nu\nu'\omega}\right],
\end{align}
where we define (in the above implicit way)  $\overline{G}$, $\widetilde{G}$ and $\widehat{{G}}$ also in the Matsubara frequency representation. We used here particle-hole frequency parametrization $\nu, \nu', \omega$ for the two-particle Green's function, i.e.
\begin{align}
&G_{1234}(\tau_1,\tau_2,\tau_3,\tau_4) \nonumber \\
&= \frac{1}{\beta^3} \sum_{\nu,\nu',\omega} G_{1234}^{\nu\nu'\omega} e^{-i\nu\tau_1} e^{i(\nu+\omega)\tau_2} e^{-i(\nu'+\omega)\tau_3} e^{i\nu'\tau_4}.
\end{align}

\section{Simplified expressions with use of the three-point correlators}
\label{sec:AppB}
Alternatively to the expression in terms of one- and two-particle Green's functions, some correlators $\cal C$ can be expressed using the Fourier transform  $g^{kq}_{d/m}$  of the three-point fermion-boson correlation functions $g_{ijl}=\langle c_i c^{\dagger}_j n_l\rangle$
in the density ($d$) or magnetic ($m$) channel  
(full definition below).
The expressions for $\cal C$ in terms of the three-point fucntions  are given by
\begin{align}
{\cal C}_{8A} &= \sum_{k,k'}G^kG^{k'}e^{i\vec{k}{(\vec{r}_i-\vec{r}_j)}}
\nonumber \\
& - \frac{1}{2}\sum_{k,q} (g^{kq}_d - g^{kq}_m + \beta G^k \langle n \rangle\delta_q)\;e^{i\vec{k}{(\vec{r}_i-\vec{r}_j)}} \nonumber \\
&= -(1-\langle n_{i\uparrow}\rangle)\sum_{k}G^ke^{i\vec{k}{(\vec{r}_i-\vec{r}_j)}}
\nonumber \\
& - \frac{1}{2U}\sum_{k,q} \left( \gamma^{kq}_d W^q_d + \gamma^{kq}_m W^q_m\right)G^kG^{k+q}e^{i\vec{k}{(\vec{r}_i-\vec{r}_j)}}
, 
\end{align}
\begin{align}
{\cal C}_{8B} &= \sum_{k,k'}G^kG^{k'}e^{i\vec{k}{(\vec{r}_i-\vec{r}_j)}}
\nonumber \\
& - \frac{1}{2}\sum_{k,q} (g^{kq}_d - g^{kq}_m + \beta G^k \langle n \rangle\delta_q)\;e^{i(\vec{k}+\vec{q}){(\vec{r}_i-\vec{r}_j)}}\nonumber \\
&= -(1-\langle n_{i\uparrow}\rangle)\sum_{k}G^ke^{i\vec{k}{(\vec{r}_i-\vec{r}_j)}}
\nonumber \\
& - \frac{1}{2U}\sum_{k,q} \left( \gamma^{kq}_d W^q_d + \gamma^{kq}_m W^q_m\right)G^kG^{k+q}e^{i(\vec{k}+\vec{q}){(\vec{r}_i-\vec{r}_j)}}
, 
\end{align}
\begin{align}
{\cal C}_{12} &= -\frac{1}{t}\sum_{k,k'}G^kG^{k'}e^{i\vec{k}{(\vec{r}_i-\vec{r}_j)}}
\varepsilon_{\vec{k}} \nonumber \\
& + \frac{1}{2t}\sum_{k,q} (g^{kq}_d - g^{kq}_m + \beta G^k \langle n \rangle\delta_q)\;e^{i\vec{k}{(\vec{r}_i-\vec{r}_j)}}
\varepsilon_{\vec{k}}, \nonumber \\
&= (1-\langle n_{i\uparrow}\rangle)\frac{1}{t}\sum_{k}G^ke^{i\vec{k}{(\vec{r}_i-\vec{r}_j)}}
\varepsilon_{\vec{k}} \nonumber \\
& + \frac{1}{2tU}\sum_{k,q} \left( \gamma^{kq}_d W^q_d + \gamma^{kq}_m W^q_m\right)G^kG^{k+q}e^{i\vec{k}{(\vec{r}_i-\vec{r}_j)}}
\varepsilon_{\vec{k}}, 
\end{align}
\begin{align}
{\cal C}_{\mu t1} &= -\frac{1}{t}\sum_{k,k'}G^kG^{k'}\varepsilon_{\vec{k}} +\frac{1}{t}\sum_{k,q}G^kG^{k+q}e^{i\vec{q}{(\vec{r}_i-\vec{r}_j)}}
\varepsilon_{\vec{k}} \nonumber \\
& + \frac{1}{t}\sum_{k,q} \left(\frac{1}{2}(g^{kq}_d + g^{kq}_m + \beta G^k \langle n \rangle\delta_q) \right.\nonumber \\ 
& \left. -G^kG^{k+q}\right)\;e^{i\vec{q}{(\vec{r}_i-\vec{r}_j)}}
\varepsilon_{\vec{k}}= (1-\langle n_{i\uparrow}\rangle)\frac{1}{t}\sum_{k}G^k\varepsilon_{\vec{k}} \nonumber \\
&+\frac{1}{t}\sum_{k,q}G^kG^{k+q}e^{i\vec{q}{(\vec{r}_i-\vec{r}_j)}}
\varepsilon_{\vec{k}} \nonumber \\
& + \frac{1}{2tU}\sum_{k,q} \left( \gamma^{kq}_d W^q_d - \gamma^{kq}_m W^q_m-2U\right)\nonumber \\
& \times G^kG^{k+q}e^{i\vec{q}{(\vec{r}_i-\vec{r}_j)}}
\varepsilon_{\vec{k}},
\end{align}
\begin{align}
{\cal C}_{\mu t2} &= -\frac{1}{t}\sum_{k,k'}G^kG^{k'}\varepsilon_{\vec{k'}} +\frac{1}{t}\sum_{k,q}G^kG^{k+q}e^{i\vec{q}{(\vec{r}_i-\vec{r}_j)}}
\varepsilon_{\vec{k}+\vec{q}} \nonumber \\
& + \frac{1}{t}\sum_{k,q} \left(\frac{1}{2}(g^{kq}_d + g^{kq}_m + \beta G^k \langle n \rangle\delta_q) \right.\nonumber \\ 
& \left. -G^kG^{k+q}\right)\;e^{i\vec{q}{(\vec{r}_i-\vec{r}_j)}}
\varepsilon_{\vec{k}+\vec{q}}= (1-\langle n_{i\uparrow}\rangle)\frac{1}{t}\sum_{k}G^k\varepsilon_{\vec{k}} \nonumber \\ 
&+\frac{1}{t}\sum_{k,q}G^kG^{k+q}e^{i\vec{q}{(\vec{r}_i-\vec{r}_j)}}
\varepsilon_{\vec{k}+\vec{q}} \nonumber \\
& + \frac{1}{2tU}\sum_{k,q} \left( \gamma^{kq}_d W^q_d - \gamma^{kq}_m W^q_m - 2U\right)\nonumber \\
&\times G^kG^{k+q}e^{i\vec{q}{(\vec{r}_i-\vec{r}_j)}}
\varepsilon_{\vec{k}+\vec{q}},
\end{align}
\begin{align}
{\cal C}_{4} &= -\sum_{k,k'}i\nu(G^k - \frac{1}{i\nu})G^{k'}e^{i\vec{k}{(\vec{r}_i-\vec{r}_j)}}
 \nonumber \\
& + \frac{1}{2}\sum_{k,q}i\nu (g^{kq}_d - g^{kq}_m + \beta G^k \langle n \rangle\delta_q)\;e^{i\vec{k}{(\vec{r}_i-\vec{r}_j)}} \nonumber \\
& = (1-\langle n_{i\uparrow}\rangle)\sum_{k}i\nu(G^k - \frac{1}{i\nu})e^{i\vec{k}{(\vec{r}_i-\vec{r}_j)}}
 \nonumber \\
& + \frac{1}{2U}\sum_{k,q}i\nu \left( \gamma^{kq}_d W^q_d + \gamma^{kq}_m W^q_m\right)G^kG^{k+q}e^{i\vec{k}{(\vec{r}_i-\vec{r}_j)}}, 
\end{align}
\begin{align}
{\cal C}_{\mu 1} &= -\sum_{k,k'}i\nu(G^k - \frac{1}{i\nu})G^{k'}  \nonumber \\
& +\sum_{k,q}i(\nu+\omega)\left(G^{k+q} - \frac{1}{i(\nu+\omega)}\right)G^{k}e^{i\vec{q}{(\vec{r}_i-\vec{r}_j)}}
 \nonumber \\
& + \sum_{k,q}i(\nu+\omega) \left( \frac{1}{2}(g^{kq}_d + g^{kq}_m + \beta G^k \langle n \rangle\delta_q) \right.\nonumber \\ 
& \left. -G^kG^{k+q}\right)e^{i\vec{q}{(\vec{r}_i-\vec{r}_j)}}= (1-\langle n_{i\uparrow}\rangle)\sum_{k}i\nu(G^k - \frac{1}{i\nu})  \nonumber \\
& +\sum_{k,q}i(\nu+\omega)\left(G^{k+q} - \frac{1}{i(\nu+\omega)}\right)G^{k}e^{i\vec{q}{(\vec{r}_i-\vec{r}_j)}}\nonumber \\
& +  \frac{1}{2U}\sum_{k,q}i(\nu+\omega) \left(\gamma^{kq}_d W^q_d - \gamma^{kq}_m W^q_m -2U\right)\nonumber \\
&\times G^kG^{k+q}e^{i\vec{q}{(\vec{r}_i-\vec{r}_j)}},
\end{align}
\begin{align}
{\cal C}_{\mu 2} &= -\sum_{k,k'}i\nu(G^k - \frac{1}{i\nu})G^{k'}  +\sum_{k,q}i\nu(G^k - \frac{1}{i\nu})G^{k+q}e^{i\vec{q}{(\vec{r}_i-\vec{r}_j)}}
 \nonumber \\
& + \sum_{k,q}i\nu \left( \frac{1}{2}(g^{kq}_d + g^{kq}_m + \beta G^k \langle n \rangle\delta_q)\right.\nonumber \\ 
& \left.  -G^kG^{k+q}\right)e^{i\vec{q}{(\vec{r}_i-\vec{r}_j)}}= (1-\langle n_{i\uparrow}\rangle)\sum_{k}i\nu(G^k - \frac{1}{i\nu}) \nonumber \\ 
& +\sum_{k,q}i\nu(G^k - \frac{1}{i\nu})G^{k+q}e^{i\vec{q}{(\vec{r}_i-\vec{r}_j)}}
 \nonumber \\
 & + \frac{1}{2U}\sum_{k,q}i\nu \left( \gamma^{kq}_d W^q_d - \gamma^{kq}_m W^q_m -2U\right)G^kG^{k+q}e^{i\vec{q}{(\vec{r}_i-\vec{r}_j)}}, 
\end{align}
where  $\langle n \rangle  = \langle n_{i\uparrow} + n_{i\downarrow} \rangle = 2\langle n_{i\uparrow} \rangle $ and $g^{kq}_{\alpha}$ is the three-point fermion-boson correlation function in the channel $\alpha=d,m$: 
\begin{align}
&g^{kq}_m = \sum_{i,j,l}\int_0^\beta \! \! \! d\tau_1 \int_0^\beta \!\! \! d\tau_2 \; e^{i\nu\tau_1} e^{-i(\nu+\omega)\tau_2} \nonumber
 \\
&\times e^{i\vec{k}(\vec{r}_i-\vec{r}_l)}e^{-i(\vec{k}+\vec{q})(\vec{r}_j -\vec{r}_l) } 
 \langle c_{i\uparrow}(\tau_1)c^\dagger_{j\uparrow}(\tau_2)S^z_l\rangle 
 \\
&g^{kq}_d = \sum_{i,j,l}\int_0^\beta \! \! \! d\tau_1 \int_0^\beta \!\! \! d\tau_2 \; e^{i\nu\tau_1} e^{-i(\nu+\omega)\tau_2} \nonumber
 \\
&\times e^{i\vec{k}(\vec{r}_i-\vec{r}_l)}e^{-i(\vec{k}+\vec{q})(\vec{r}_j -\vec{r}_l) } 
 \langle c_{i\uparrow}(\tau_1)c^\dagger_{j\uparrow}(\tau_2)n_l\rangle 
\end{align}
with $S^z_i = n_{i\uparrow}-n_{i\downarrow}$ and $n_i = n_{i\uparrow}+n_{i\downarrow}$. 

The three-point correlator is related to the fermion-boson irreducible vertex $\gamma^{kq}_\alpha$ as follows~\cite{Krien:PRB19}:
\begin{equation}
g^{kq}_\alpha = \gamma^{kq}_\alpha\frac{W^q_\alpha}{U_\alpha}G^kG^{k+q}  -  \beta G^k \langle n \rangle\delta_q \delta_{\alpha,d},
\end{equation}
where $W^q_\alpha$ is related to the respective susceptibility $\chi^q_\alpha$ in the following way~\cite{Krien:PRB19,Krien2020b}:
\begin{equation}
 W^q_\alpha = U_\alpha -  \frac{1}{2}U_\alpha\chi^q_\alpha U_\alpha,
 \label{eq:Wdef}
 \end{equation}
 with $U_d = U$ and $U_m =-U$.

 For completeness we also give below the explicit expressions for susceptibilities:
\begin{align}
&\chi^q_m = \sum_{j}\int_0^\beta \! \! d\tau  \;e^{-i\omega \tau} e^{-i\vec{q}{(\vec{r}_i-\vec{r}_j)}}\langle S^z_i(\tau)S^z_j\rangle \\ 
&\chi^q_d = \sum_{j}\int_0^\beta \! \! d\tau e^{-i\omega \tau} e^{-i\vec{q}{(\vec{r}_i-\vec{r}_j)}}\langle n_i(\tau)n_j\rangle -\beta \delta_q \langle n_i\rangle^2
\\ 
&\chi^q_s = \sum_{j}\int_0^\beta \! \! d\tau e^{-i\omega \tau} e^{-i\vec{q}{(\vec{r}_i-\vec{r}_j)}}\langle c_{i\downarrow}(\tau)c_{i\uparrow}(\tau)c^\dagger_{j\uparrow}c^\dagger_{j\downarrow}\rangle.
\end{align}

The remaining $\cal{C}$ correlators cannot be expressed  by three-point correlation functions, they require the full four-point (two-particle) vertex $F_{\sigma\sigma'}^{kk'q}$, defined as  follows~\cite{RohringerRMP}:
\begin{align}
{G}_{\sigma\sigma'}^{kk'q} &= G^k G^{k'}\delta_{q0} - G^k G^{k+q}\delta_{kk'}\delta_{\sigma\sigma'} \nonumber \\
&- G^k G^{k+q}{F}_{\sigma\sigma'}^{kk'q}G^{k'}G^{k'+q} 
\end{align}
In terms of $F_{\sigma\sigma'}^{kk'q}$, we obtain the following expressions:
	\begin{align}
	 &{\cal C}_1 = -\sum_{k}\nu\left( G^k - \frac{1}{i\nu} \right)\sum_{k'}\nu'\left(G^{k'} - \frac{1}{i\nu'} \right)\nonumber \\
  &+\sum_{k,q}\nu(\nu+\omega)\left(G^k-  \frac{1}{i\nu}\right)\left(G^{k+q} - \frac{1}{i(\nu+\omega)} \right)e^{i\vec{q}{(\vec{r}_i-\vec{r}_j)}}\nonumber \\
  &+ \frac{1}{2}\sum_{k,k',q}\nu(\nu'+\omega)G^k G^{k+q}G^{k'}G^{k'+q} \nonumber \\
  &\times (F_d^{kk'q}+ F_m^{kk'q})e^{i\vec{q}{(\vec{r}_i-\vec{r}_j)}}, 
  \label{eq:C1B}
	\end{align}
	\begin{align}
	 &{\cal C}_{t1} = \frac{1}{t}\sum_{k} \varepsilon_{\vec{k}}  G^k \sum_{k'}i\nu'\left(G^{k'} - \frac{1}{i\nu'} \right)\nonumber \\
  &- \frac{1}{t}\sum_{k,q}i(\nu+\omega)G^k \left(G^{k+q} - \frac{1}{i(\nu+\omega)} \right)e^{i\vec{q}{(\vec{r}_i-\vec{r}_j)}}\varepsilon_{\vec{k}}   \nonumber \\
  &- \frac{1}{2t}\sum_{k,k',q}i(\nu'+\omega)G^k G^{k+q}G^{k'}G^{k'+q} \nonumber \\
  &\times (F_d^{kk'q}+ F_m^{kk'q})e^{i\vec{q}{(\vec{r}_i-\vec{r}_j)}}\varepsilon_{\vec{k}}, 
	\end{align}
 	\begin{align}
	 &{\cal C}_{t2} = \frac{1}{t}\sum_{k}i\nu\left(G^{k} - \frac{1}{i\nu} \right)  \sum_{k'}\varepsilon_{\vec{k'}}G^{k'} \nonumber \\
  &- \frac{1}{t}\sum_{k,q}i\nu\left(G^{k} - \frac{1}{i\nu} \right)G^{k+q} e^{i\vec{q}{(\vec{r}_i-\vec{r}_j)}}\varepsilon_{\vec{k}+\vec{q}}   \nonumber \\
  &- \frac{1}{2t}\sum_{k,k',q}i\nu G^k G^{k+q}G^{k'}G^{k'+q} \nonumber \\
  &\times (F_d^{kk'q}+ F_m^{kk'q})e^{i\vec{q}{(\vec{r}_i-\vec{r}_j)}}\varepsilon_{\vec{k}'+\vec{q}}, 
	\end{align}

  	\begin{align}
	 &{\cal C}_{t^2} = \frac{1}{t^2}\sum_{k}\varepsilon_{\vec{k}}G^{k} \sum_{k'}\varepsilon_{\vec{k'}}G^{k'} \nonumber \\
  &- \frac{1}{t^2}\sum_{k,q}G^{k}G^{k+q} e^{i\vec{q}{(\vec{r}_i-\vec{r}_j)}}\varepsilon_{\vec{k}}\varepsilon_{\vec{k}+\vec{q}}   \nonumber \\
  &- \frac{1}{2t^2}\sum_{k,k',q}G^k G^{k+q}G^{k'}G^{k'+q} \nonumber \\
  &\times (F_d^{kk'q}+ F_m^{kk'q})e^{i\vec{q}{(\vec{r}_i-\vec{r}_j)}}\varepsilon_{\vec{k}}\varepsilon_{\vec{k}'+\vec{q}}.
	\end{align}
Please note, that knowing the four-point vertex is  sufficient to calculate all other $\cal{C}$ correlators. In particular, the three-point fermion-boson vertex is given by~\cite{Krien2020b}
\begin{align}
  &\gamma^{kq}_\alpha = \frac{1 + \sum_{k'}
   F_\alpha^{kk'q}G^{k'}G^{k'+q} }{1-\frac{1}{2}U_\alpha\chi_\alpha^q}
   \label{eq:gamma_def}
\end{align}
and the susceptibility by
\begin{align}
  &\chi_\alpha^q = -2\sum_k G^k G^{k+q} -2\sum_{k,k'}G^k G^{k+q}F_\alpha^{kk'q}G^{k'}G^{k'+q}.
\end{align}

\section{Convergence with respect to the number of Matsubara frequencies}
\label{App:convergence}
The Matsubara summations in Eqs.~\eqref{eq:C1}-\eqref{eq:C13} run over an infinite number of frequencies. Typical  computations with two-particle Green's functions or vertices use finite, often even relatively  small frequency boxes. This can be remedied, at least in part, by properly taking into account the asymptotic behavior.

For the one-particle Green's function the asymptotic behavior at large frequencies is known and can be easily  accounted for. For vertex functions computing the asymptotic behavior is more involved~\cite{Wentzell20, Krien2022} and requires knowledge of two- and three-point correlation functions that depend on one and two Matsubara frequencies respectively. These functions are computationally less demanding than the four-point vertices and thus can be obtained within a bigger frequency range. 

In the following we will show the frequency box-size dependence of parts of the ${\cal C}_1$ correlator defined in Eq.~\eqref{eq:C1}, but calculated from Eq.~\eqref{eq:C1B}. We will treat separately the parts that can be obtained from one-particle Green's function only ('bubble') and from the two-particle ('vertex') functions.

To compute the first two terms of~\eqref{eq:C1B} we need the one-particle Green's function $G^k$. We can divide $G^k$ in its asymptotic part and the rest $G_R^k$ in the following way
\begin{align}
G^k &= \frac{1}{i\nu - \xi_{\vec{k}}} + G^k_{R},
\label{eq:gf_as2}
\end{align}
where we defined $\xi_{\vec{k}} =\varepsilon_{\vec{k}} + \frac{1}{2}U\langle n \rangle -\mu$.
The rest part  ${G^k_R}$
 decays at least as $1/(\nu)^2$ for large frequencies, which makes frequency summations of ${G^k_R}$
well convergent. The slowly decaying parts can be summed analytically. In fact, they have to be summed analytically, because the infinite sum of $1/\nu$ is only convergent through evaluation of a contour integral -- the choice of the contour depends on which limit of the $G(\tau)$ we need. For $\tau \to 0^+$ we obtain 
\begin{align}
   \frac{1}{\beta}\sum_\nu \frac{1}{i\nu-\xi_{\vec{k}}} e^{-i\nu0^+}&=  -\frac{1}{1+e^{-\beta\xi_{\vec{k}}}},\\
    \sum_k i\nu \left (G^k - \frac{1}{i\nu} \right)e^{-i\nu0^{+}}&   = \sum_k i\nu {G}^k_{R} e^{-i\nu0^+} \nonumber \\ &- \frac{1}{N_k}\sum_{\vec{k}}\frac{\xi_{\vec{k}}}{1+e^{-\beta\xi_{\vec{k}}}} . 
\end{align}

In order to increase the accuracy of the frequency sum evaluation of the two-particle vertex contribution to~Eq.~\eqref{eq:C1B}, we separate from the vertex $F$ the single-boson exchange contribution  $\gamma_\alpha W_\alpha \gamma_\alpha$ (for details of the single-boson exchange decomposition see~\cite{Krien:PRB19,Krien2020b}) and denote the rest by $T_{\alpha}$
\begin{align}
F_{\alpha}^{kk'q} = T_{\alpha}^{kk'q} + \gamma_\alpha^{kq}W_\alpha^q \gamma_\alpha^{k'q}.
\label{eq:Tgwg}
	\end{align}
The screened interaction $W_{\alpha}$, defined in Eq.~\eqref{eq:Wdef}, and the three-point fermion-boson vertex $\gamma_\alpha$ have (i) a reduced frequency dependence as compared to $F_{\alpha}$ or $T_{\alpha}$ and can thus be computed in a much bigger frequency window and have (ii) exactly known asymptotics for large frequencies, so that the frequency sum can be extended beyond the computed box. Alternatively to the above approach, a different strategy to include vertex asymptotics can be employed, as in \cite{Wentzell20}, \cite{Li16}, or \cite{Kunes11}.

\begin{figure}[tb]
\includegraphics[width=\linewidth]{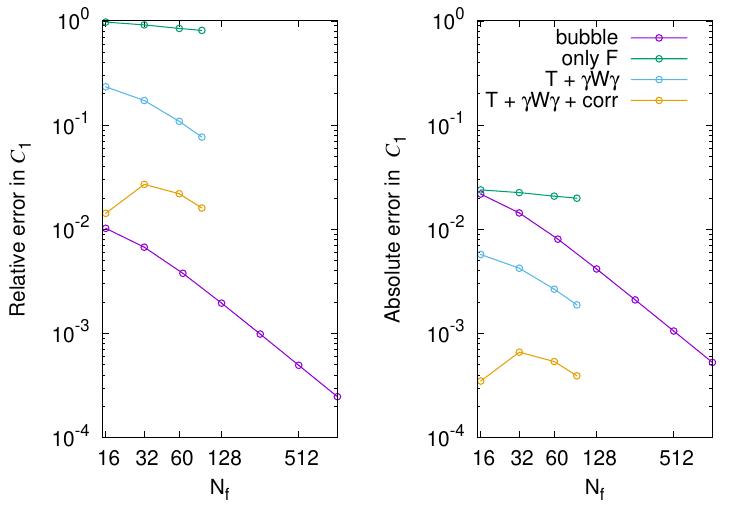}
\caption{Relative (left) and absolute (right) error as a function of the number of Matsubara frequencies $N_f$ (linear dimension of the frequency box) used in the sum in Eq.~\eqref{eq:C1B} needed to obtain the bubble and vertex contributions to the ${\cal C}_1$ correlator. The errors are computed with respect to values obtained from  the evaluation of Eq.~\eqref{eq:C1} in the imaginary time domain. Please note the log-log scale. The data were obtained from ED for nearest neighbors (NN) in the  $2\times2$ Hubbard cluster at half filling ($U=1$, $\beta=10$, $\mu = 0.5$). }
\label{fig:convergence1}
\end{figure}

In Fig.~\ref{fig:convergence1} we show the dependence of the  relative and absolute error made by summation over a finite frequency box as a function of the linear dimension of the box $N_f$. We define the absolute error as $|C^{freq}-C^{time}|$ and the relative error as $|(C^{freq}-C^{time})|/|C^{time}|$, where $C^{freq}$ is the quantity computed as a frequency sum with a given $N_f$ and $C^{time}$ is the same quantity computed directly in imaginary time domain. We use as an example the 'bubble' and 'vertex' part of the ${\cal C}_1$ correlator calculated from Eq.~\eqref{eq:C1B} with ED for the $2\times2$ Hubbard cluster at half filling at $U=1$, $\beta=10$, $\mu = 0.5$ and for the neighbouring sites. We see that both the relative and absolute  error of the bubble contribution computed using Eq.~\eqref{eq:gf_as2} quickly decay wit $N_f$. 

For the vertex part it plays a crucial role if only $F$ is used (green curves) or the asymptotic treatment of Eq.~\eqref{eq:Tgwg} is used (blue curves). Further improvement can be achieved by subtracting from $T_{\alpha}$ and adding a single-boson exchange contribution  $\gamma_\alpha^{kq}W_\alpha^q \gamma_\alpha^{k'q}$ in the transversal particle-hole channel (yellow curves, denoted as "$T+\gamma W\gamma +$corr" in the legend), i.e.,  subtracting this term from $T_{\alpha}$ in the 'inner' frequency box and adding the asymptotics-extended sum. The fact that the smallest frequency box seems to give the most accurate result for the yellow curves is accidental and not observed for other parameters. 
\begin{figure}[tb]
\includegraphics[width=\linewidth]{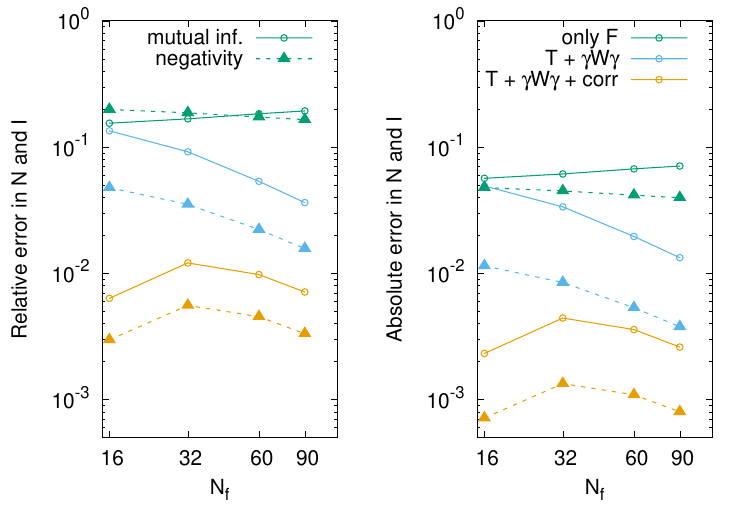}
\caption{Relative (left) and absolute (right) error in the mutual information (full lines with circles) and negativity (dashed lines with triangles) as a function of the number of Matsubara frequencies $N_f$ (linear dimension of the frequency box) used in computing the vertex contribution to the ${\cal C}_1$ correlator. The errors are computed for the same parameters and color coding as in Fig.~\ref{fig:convergence1}.} 
\label{fig:convergence2}
\end{figure}

In Fig.~\ref{fig:convergence2}, we show the influence of the error made in computing the vertex contribution to ${\cal C}_1$ on the overall result for the negativity and mutual information. The parameters and color coding are the same as in Fig.~\ref{fig:convergence1}.  It is visible that the error in the vertex part of ${\cal C}_1$ has significant influence on the overall result in our example case. Again, the convergence with frequency box size is very much improved by using the known asymptotics.  This has to be taken into account in any future computations of entanglement measures from frequency sums of vertex functions.

\section{Density matrix of a single site}
\label{Appendix1siteRD}
	To calculate the mutual information between two sites one also needs the density matrix of a single site. It is much more simple than the two-site density matrix. At first, there are only four basis vectors see Table \ref{tab:one-site}.
	\begin{table}[h!]
		\begin{center}
			\caption{Basis in the subspace of a given site $i$. }
			\label{tab:one-site}
			\begin{tabular}{|c |c c   |} 
				& i$\uparrow$ & i$\downarrow$  \\ \hline
				$v_1$   & 0  & 0 \\ \hline
				$v_2$   & 0  & 1 \\
				$v_3$   & 1  & 0 \\
				$v_4$   & 1  & 1 \\
				\end{tabular}
		\end{center}
	\end{table}
The second good news is that this 4 by 4 matrix is diagonal, particle number and spin conservation imply that all off-diagonal elements are zero.
\begin{equation}
\rho_i = \left[ \begin{array}{cccc}
	\rho_{1,1} & & & \\
	& \rho_{2,2} & & \\
	& & \rho_{3,3} &\\
	& & & \rho_{4,4}
\end{array} \right]
\end{equation}
The diagonal elements are given by the following expectation values
\begin{align}
	\rho_{1,1} &= \langle  (1-n_{i\uparrow})(1-n_{i\downarrow}) \rangle\\
	\rho_{2,2} &= \langle  n_{i\uparrow}(1-n_{i\downarrow}) \rangle\\
	\rho_{3,3} &= \langle  (1-n_{i\uparrow})n_{i\downarrow} \rangle\\
	\rho_{4,4} &= \langle  n_{i\uparrow}n_{i\downarrow} \rangle
 \; .
\end{align}
These are related to the correlators ${\cal C}_{13}$ and  ${\cal C}_7$ through Eqs.~(\ref{eq:niup}) and (\ref{eq:niupnidown}); note $\langle n_{i\uparrow}\rangle=\langle n_{i\downarrow}\rangle$ for SU(2) symmetry. 

\section{Spectrum of the reduced density matrix}
\label{AppendixEV}
Although the dimension of the reduced density matrix is 16, it has the following symmetries: particle number and spin conservation, interchanging the two sites. 
This makes possible to give closed expression for the eigenvalues $\lambda_{i=1\ldots16}$ of the two-site density matrix:
\begin{align}
\lambda_1 &=\rho_{1,1} \\ 
\lambda_2 &= \rho_{2,2} - \rho_{2,4}  \\
\lambda_3 &=  \rho_{2,2} + \rho_{2.4} \\
\lambda_4 &=  \rho_{2,2} - \rho_{3,5} \\
\lambda_5 &=  \rho_{2,2} + \rho_{3,5} \\
\lambda_6 &=\rho_{8,8} + \rho_{8,9}\\
\lambda_7 &=\rho_{7,7} \\
\lambda_{8} &=\rho_{6,6} - \rho_{6,11}\\
\lambda_9 & = \frac{1}{2} b  \nonumber\\
&+\frac{1}{2} \sqrt{b^2 + 16 \rho^2_{6,8} - 4 (\rho_{6,6}+\rho_{6,11})(\rho_{8,8}-\rho_{8,9}) } \\
\lambda_{10} & = \frac{1}{2} b  \nonumber\\
&-\frac{1}{2} \sqrt{b^2 + 16 \rho^2_{6,8} - 4 (\rho_{6,6}+\rho_{6,11})(\rho_{8,8}-\rho_{8,9}) } \\
&b=\rho_{8,8}-\rho_{8,9}+\rho_{6,6}+\rho_{6,11} \\
\lambda_{11}&=\rho_{10,10} \\
\lambda_{12} &=  \rho_{12,12} - \rho_{12,14} \\
\lambda_{13} &=  \rho_{12,12} + \rho_{12,14}\\
\lambda_{14} &=  \rho_{12,12} - \rho_{13,15}\\
\lambda_{15} &=   \rho_{12,12} + \rho_{13,15} \\
\lambda_{16} &=\rho_{16,16}
\end{align}

\bibliography{main}

\end{document}